\documentclass[sigconf, authordraft, review=false, timestamp=false]{acmart}

\newtheorem{proposition}{Proposition}
\usepackage{multirow}
\usepackage[table]{xcolor}

\usepackage{hyperref}
\usepackage{url}
\usepackage{graphicx}
\usepackage{wrapfig}
\usepackage{booktabs}
\usepackage{multirow}
\usepackage{array}
\usepackage{siunitx}
\usepackage{tabularx}
\usepackage{xcolor}
\usepackage{colortbl,booktabs}
\usepackage{enumitem}
\usepackage{pifont}
\usepackage[most]{tcolorbox}
\usepackage{hyperref}
\usepackage{arydshln}

\usepackage{algorithm}
\usepackage{algpseudocode}
\usepackage{cancel}
\usepackage{circledsteps}

\definecolor{negativered}{RGB}{255, 0, 0} 
\definecolor{positiveblue}{RGB}{30, 144, 255} 
\definecolor{rowgray}{gray}{0.851}

\newcommand{\cmark}{\ding{51}} 
\newcommand{\xmark}{\ding{55}}  

\AtBeginDocument{%
  }

\usepackage{graphicx} 
\usepackage{subcaption}

\begin{document}

\title{Towards Fair Large Language Model-based Recommender Systems without Costly Retraining}
\author{Jin Li}
\email{jin.li-4@student.uts.edu.au}
\orcid{0000-0001-5737-3594}
\affiliation{%
  \institution{University of Technology Sydney}
  \city{Sydney}
  \country{Australia}}

\author{Huilin Gu}
\email{huilin.gu@student.uts.edu.au}
\orcid{0009-0002-6116-2524}
\affiliation{%
  \institution{University of Technology Sydney}
  \city{Sydney}
  \country{Australia}
}

\author{Shoujin Wang}
\authornote{The corresponding author.}
\email{shoujin.wang@uts.edu.au}
\orcid{0000-0003-1133-9379}
\affiliation{%
  \institution{University of Technology Sydney}
  \city{Sydney}
  \country{Australia}
}

\author{Qi Zhang}
\email{zhangqi_cs@tongji.edu.cn}
\orcid{0000-0002-1037-1361}
\affiliation{%
  \institution{Tongji University}
  \city{Shanghai}
  \country{China}
}

\author{Shui Yu}
\email{shui.yu@uts.edu.au}
\orcid{0000-0003-4485-6743}
\affiliation{%
  \institution{University of Technology Sydney}
  \city{Sydney}
  \country{Australia}
}

\author{Chen Wang}
\email{chen.wang@data61.csiro.au}
\orcid{0000-0002-3119-4763}
\affiliation{%
  \institution{CSIRO Data61}
  \city{Sydney}
  \country{Australia}
}

\author{Xiwei Xu}
\email{xiwei.xu@data61.csiro.au}
\orcid{0000-0002-2273-1862}
\affiliation{%
  \institution{CSIRO Data61}
  \city{Sydney}
  \country{Australia}
}

\author{Fang Chen}
\email{fang.chen@uts.edu.au}
\orcid{0000-0003-4971-8729}
\affiliation{%
  \institution{University of Technology Sydney}
  \city{Sydney}
  \country{Australia}
}

\renewcommand{\shortauthors}{Jin Li et al.}

\begin{abstract}
Large Language Models (LLMs) have revolutionized Recommender Systems (RS) through advanced generative user modeling. However, LLM-based RS (LLM-RS) often inadvertently perpetuates bias present in the \textbf{training data}, leading to severe fairness issues. Addressing these fairness problems in LLM-RS faces two significant challenges. 1) Existing debiasing methods, designed for specific bias types, \textbf{lack the generality} to handle diverse or emerging biases in real-world applications. 2) Debiasing methods relying on retraining are \textbf{computationally infeasible} given the massive parameter scale of LLMs. To overcome these challenges, we propose \textbf{FUDLR} (Fast Unified Debiasing for LLM-RS). The core idea is to reformulate the debiasing problem as an efficient machine unlearning task with two stages. First, FUDLR identifies bias-inducing samples to unlearn through a novel bias-agnostic mask, optimized to balance fairness improvement with accuracy preservation. Its bias-agnostic design allows adaptability to various or co-existing biases simply by incorporating different fairness metrics. Second, FUDLR performs efficient debiasing by estimating and removing the influence of identified samples on model parameters. Extensive experiments demonstrate that FUDLR effectively and efficiently improves fairness while preserving recommendation accuracy, offering a practical path toward socially responsible LLM-RS. The code and data are available at \url{https://github.com/JinLi-i/FUDLR}.
\end{abstract}




\maketitle

\begin{figure}
  \centering
  \includegraphics[width=0.47\textwidth]{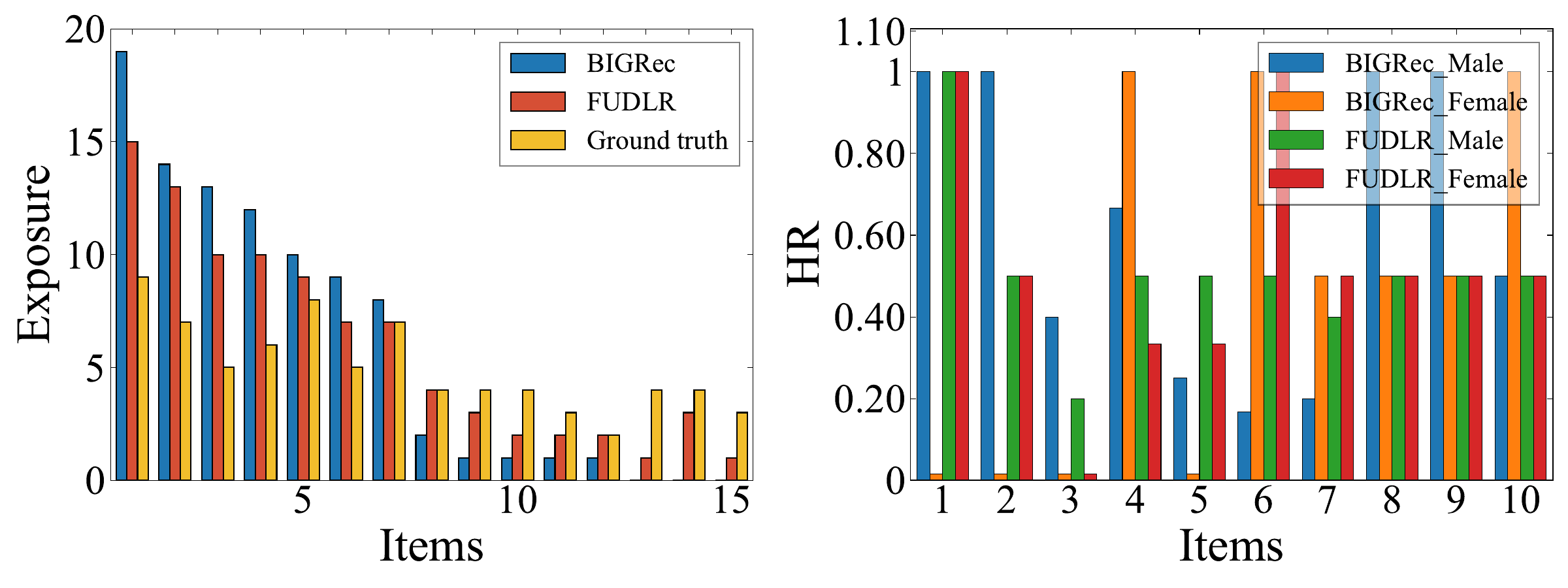}
  \vspace{-10pt}
  \caption{Observations in the ML1M dataset.
  (a) Popularity bias: the backbone recommender (e.g., BIGRec \cite{DBLP:journals/tors/BaoZWZYLCFT25}) exhibits a clear tendency to over-expose popular items while under-exposing long-tail items. Our FUDLR framework substantially alleviates this bias and produces a more balanced recommendation distribution aligned with true user preferences. (b) Attribute bias: our proposed FUDLR framework markedly improves fairness by reducing gender-related HR disparities, while the backbone model (e.g., BIGRec \cite{DBLP:journals/tors/BaoZWZYLCFT25}) displays a significant performance gap between user groups. 
  }
  \label{fig:Figure_1_2}
  \vspace{-10pt}
\end{figure}

\section{Introduction}
Recommender Systems (RS) have become an integral part of modern web and social media platforms, providing users with personalized content and suggestions and reshaping users' online experience. Recently, the emergence of Large Language Models (LLMs) \cite{Naveed2025Comprehensive,DBLP:journals/air/Kumar24,chang2024survey,grosse2023studying,guo2024large} has shown promising capabilities in understanding and generating high-quality response for various tasks \cite{DBLP:journals/corr/abs-2509-17784,li2025survey}. This motivates the integration of LLMs into recommendation tasks, leading to a paradigm shift towards generative LLM-based Recommender Systems (LLM-RS) \cite{zhou2025large,DBLP:journals/tkde/QuFZL25,DBLP:conf/ecir/JiLXHGTZ24,liu2023llmrec,DBLP:journals/tors/BaoZWZYLCFT25}. Despite their promising performance, LLM-RS are susceptible to inheriting biases present in the training data \cite{DBLP:conf/bigdataconf/SakibD24,DBLP:conf/recsys/Tommasel24}. This leads to serious, substantial fairness concerns and poses critical societal risks in online platforms.

Recent studies evidence that LLM-RS is vulnerable to various types of bias, such as popularity bias (where popular items are over-recommended) and attribute bias (where certain user groups are discriminated), as illustrated in Figure \ref{fig:Figure_1_2} (a) and (b) respectively, potentially harming the multi-stakeholders in the recommendation ecosystem. For instance, Jiang et al. \cite{DBLP:conf/www/JiangBZW0F024} identify the pattern of popularity bias in LLM-RS, where popular items are over-recommended, leading to unfair treatment for niche items. Similarly, a series of studies \cite{Deldjoo2025CFaiRLLM,DBLP:conf/recsys/ZhangBZWF023,DBLP:conf/recsys/Tommasel24,DBLP:journals/ipm/ShenLBMS23,DBLP:conf/eacl/HuaGXJLZ24} examine the user-side concerns and reveal that LLM-RS, alongside the intrinsic stereotype bias in LLMs \cite{DBLP:journals/corr/abs-2504-04199} to certain groups of users, shows discriminatory behavior against users' sensitive attributes like gender and race.

Despite the growing awareness of fairness issues in LLM-RS, addressing these identified biases remains largely underexplored and presents unique challenges (CH) compared with traditional RS.
\textbf{CH1: Lack of Generality.} Existing debiasing methods for LLM-RS, which are typically designed for a specific type of bias, have demonstrated effectiveness \cite{DBLP:conf/www/JiangBZW0F024,Deldjoo2025CFaiRLLM,DBLP:conf/recsys/ZhangBZWF023,DBLP:conf/recsys/Tommasel24}. However, they lack generality and flexibility needed to adapt to various bias types. This limitation is particularly critical in LLM-RS, as models are often trained on massive, unaligned datasets, which can introduce newly emerging biases or multiple co-existing types of biases. Meanwhile, users usually have diverse fairness requirements, necessitating adaptable debiasing solutions. Therefore, a unified debiasing framework that can effectively mitigate various types of biases in LLM-RS is highly desirable.
\textbf{CH2: Computational Constraints of LLMs.} Most existing work for debiasing LLM-RS uses reweighting \cite{DBLP:conf/www/JiangBZW0F024,DBLP:conf/www/GaoCYHY025}, data augmentation \cite{wang-etal-2023-improving-conversational}, or adversarial learning \cite{DBLP:conf/eacl/HuaGXJLZ24} approaches, and thus requires costly full model retraining or fine-tuning. Retraining from scratch is computationally prohibitive, and fine-tuning for each scenario significantly increases the operational cost for dynamic, large-scale recommender systems in practice.

To address these limitations, this paper introduces \textbf{FUDLR}, a Fast and Unified Debiasing framework for LLM-based Recommenders. FUDLR reformulates the debiasing task from a novel machine unlearning perspective, performing targeted bias mitigation without expensive retraining. It operates via a two-stage process: 1) \textbf{Bias Identification} and 2) \textbf{Fast Debiasing via Unlearning}. Specifically, FUDLR first identifies the bias-inducing training samples with a novel mask learning mechanism. This mask is optimized by jointly balancing three objectives: maximizing fairness improvement, preserving recommendation accuracy, and ensuring the intervention (the subset of biased training samples selected for removal) taken for debiasing is sparse and targeted. The core of this stage is its bias-agnostic design. By adopting a differentiable metric that quantifies a specific bias (e.g., popularity, demographic parity), the framework can be directed to mitigate it without any other algorithmic changes. Based on the bias-inducing samples identified by the learned mask, FUDLR then performs efficient debiasing via a fast unlearning update. By estimating the influence of the identified biased samples on the model parameters, FUDLR computes a one-step parameter correction to effectively remove their biasing effect. Built upon these two stages, FUDLR provides a general, efficient and practical solution for debiasing LLM-RS without costly retraining.

Our main contributions are summarized as follows:
\begin{itemize}[left=0pt]
  \item We propose FUDLR, a general and efficient framework specialized for debiasing LLM-based recommender systems. FUDLR effectively addresses various biases through a unified approach while being computationally practical for large-scale LLMs.
  \item We introduce a novel mask learning mechanism to precisely identify bias-inducing training data. Its bias-agnostic design enables the mitigation of various types of bias by substituting only the fairness metric corresponding to each type of bias.
  \item We devise a fast unlearning update to correct the model by estimating and removing the influence of identified bias-inducing samples on model parameters, circumventing the need for costly retraining and making debiasing practical for large-scale LLMs.
  \item Extensive experimental results on real-world datasets demonstrate that FUDLR effectively mitigates both item-side popularity bias and user-side attribute bias, achieving a superior balance of fairness and accuracy compared to the state-of-the-art debiasing baselines for LLM-RS.
\end{itemize}

\begin{figure}
  \centering
  \includegraphics[width=0.47\textwidth]{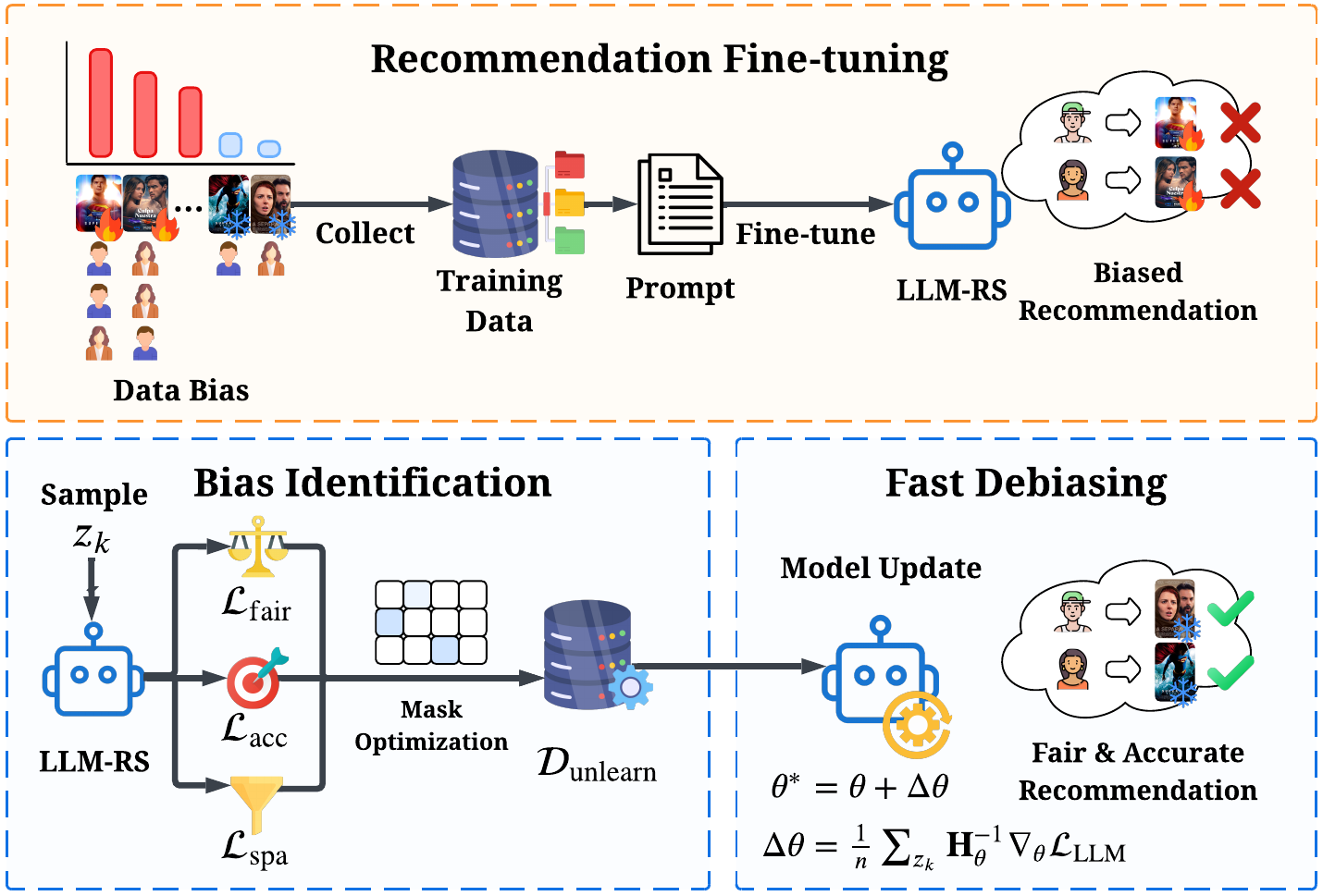}
  \caption{The framework of FUDLR. For the fine-tuned LLM-RS model, it first identifies the bias-inducing training samples via a novel mask learning mechanism, which optimizes a flexible objective balancing fairness and accuracy. Then, it performs efficient debiasing by estimating and removing the influence of the identified samples on the model parameters.}
  \label{fig:framework}
  \vspace{-15pt}
\end{figure}

\section{Related Work}
\subsection{LLM-based Recommender Systems}
To leverage the powerful language understanding and generation capabilities of LLMs in recommendation tasks, recent works mainly fall into two categories: (1) \textbf{LLM as Representation Learners}~\cite{DBLP:conf/www/ZhaoZZYLJJG25,He2025Llm2rec,Qu2025Tokenrec,ren2024representation, Lin2025Order,hu2024enhancing} and (2) \textbf{LLM as Generative Recommenders}~\cite{DBLP:journals/tors/BaoZWZYLCFT25,Zheng2024Adapting,Zhu2024Collaborative,Yin2025Unleash,nie2024hybrid,huang2025recommender}. 

The first category includes methods utilizing LLMs to extract rich user and item representations \cite{Liu2025Llmemb,Liu2024Llm,Jia2025Learn} that are then fed into traditional recommendation models for improved prediction performance. In this way, the semantic knowledge captured by LLMs can be injected into conventional recommendation frameworks. For instance, Hu et al. \cite{He2025Llm2rec} propose LLM2Rec, which integrates collaborative filtering signals with LLM-derived semantic embeddings to enhance in-domain and out-of-domain recommendation performance. One step further, Lin et al. \cite{Lin2025Order} design order-agnostic identifiers to reduce token dependency, thereby mitigating the local optima issue and performing parallel generation for efficiency.

The second category directly fine-tunes LLMs and prompts them for recommendations in a generative manner \cite{DBLP:conf/coling/LiZLC24,DBLP:journals/tkde/QuFZL25,DBLP:conf/ecir/JiLXHGTZ24,DBLP:journals/corr/abs-2507-06507,DBLP:journals/tors/BaoZWZYLCFT25}. One pioneering work is TALLRec \cite{Bao2023Tallrec}, which presents an effective and efficient fine-tuning framework by using LoRA \cite{Hu2022Lora} to adapt LLMs for recommendation tasks. However, the intrinsic non-deterministic nature of LLMs poses challenges for recommending aligned items. To address this, Bao et al. \cite{DBLP:journals/tors/BaoZWZYLCFT25} propose a two-step grounding framework that first aligns LLMs to the recommendation space by fine-tuning on user-item interaction histories, followed by a grounding step that matches generated candidates to actual items. 

Our work focuses on the paradigm of LLM being Generative Recommenders, as it fully leverages the generative power of LLMs for user modeling and recommendation and it faces more severe fairness challenges \cite{DBLP:conf/www/JiangBZW0F024,Deldjoo2025CFaiRLLM,DBLP:conf/recsys/ZhangBZWF023,DBLP:conf/recsys/Tommasel24} and requires specifically tailored alignment and debiasing solutions.

\subsection{Fairness in LLM-RS}
While fairness in traditional recommender systems has been extensively studied \cite{Wang2023Survey,DBLP:journals/umuai/DeldjooJBDZ24,DBLP:journals/tist/LiCXGTLZ23,song2023counterfactual,ge2022explainable,wang2024hierarchical,DBLP:journals/tist/WangZWR24, fu2020fairness,DBLP:journals/corr/abs-2402-08241,DBLP:conf/www/LiWZYC25}, fairness in LLM-based Recommender Systems is an emerging area of research. Recent studies have identified various biases in LLM-RS. Acknowledging that RS ecosystems involve multiple stakeholders, we examine the representative empirical studies and existing debiasing methods for LLM-RS from both the item and user perspectives.

\textbf{Item-side Fairness} mainly focuses on the equality of different groups of items being exposed and recommended. One of the most representative biases on this side is \textbf{popularity bias}, where popular items are over-recommended, leading to unfair treatment for niche items. As identified in \cite{DBLP:conf/www/JiangBZW0F024}, LLM-RS can exacerbate popularity bias due to their tendency to generate frequent patterns seen during training. To mitigate this, Jiang et al. \cite{DBLP:conf/www/JiangBZW0F024} propose a reweighting strategy used in the retraining process, where the loss contribution of each sample is rescaled based on group-level interaction proportions. Meanwhile, a reranking strategy is also designed as a post-processing step to rerank the items of different groups by introducing an additional punishment term. In the field of conversational recommendation, strategies of data augmentation \cite{wang-etal-2023-improving-conversational} and prompt tuning \cite{DBLP:journals/corr/abs-2401-03605} have also been explored for debiasing.

\textbf{User-side Fairness} aims to ensure equitable treatment across different user groups, often defined by sensitive attributes, such as gender, age, and occupation. Existing research \cite{DBLP:conf/recsys/ZhangBZWF023,Deldjoo2025CFaiRLLM,DBLP:conf/recsys/Tommasel24} has found that when LLMs are exposed with sensitive user attributes, they tend to generate recommendations that reinforce existing stereotypes \cite{DBLP:journals/corr/abs-2504-04199} and discriminate against certain user groups. A straightforward solution is to perform prompt masking \cite{DBLP:journals/ipm/ShenLBMS23} to remove sensitive attributes from the input prompts. However, this strategy does not essentially correct the model and it falls short for implicit attribute bias where attribute patterns are embedded in behavioral data. To tackle this scenario, Hua et al. \cite{DBLP:conf/eacl/HuaGXJLZ24} propose a Counterfactually-Fair-Prompt method trained by adversarial learning to remove sensitive information in the user token embeddings.

Despite the growing awareness of fairness issues in LLM-RS, the critical challenges persist: 1) The lack of generality in existing debiasing methods limits their applicability to diverse or emerging biases. 2) The computational constraints of large-scale LLMs hinder the practicality of current training-based debiasing approaches. Our proposed FUDLR framework addresses these challenges by providing a unified and efficient debiasing approach that can adapt to various bias types without costly retraining.

\section{Problem Formulation}
\label{sec:problem_formulation}
Let $\mathcal{U} = \{u_1, u_2, \cdots, u_M\}$ and $\mathcal{I} = \{i_1, i_2, \cdots, i_N\}$ denote the sets of users and items (e.g., the name of movies), respectively. The sequence $\mathcal{S}^u_{1:n} 
= [i_1^u, i^u_2, \cdots, i^u_n]$ records the items that user $u$ has interacted with. Generally, the goal of a sequential recommender system is to predict the item $i_{n+1}^u$ that user $u$ interacted with at the $(n+1)$-th timestamp. Formally, $\hat{i}_{n+1}^u = \arg\max_{i \in \mathcal{I}} P(i | \mathcal{S}^u_{1:n}; \theta)$, where $\theta$ denotes the model parameters. 

\textbf{LLM-based Recommendation.} Recently, inspired by the success of LLMs in various NLP tasks, numerous studies have explored the potential of LLMs for recommendation tasks \cite{DBLP:journals/tors/BaoZWZYLCFT25,Zheng2024Adapting,Zhu2024Collaborative,Yin2025Unleash}. In this paper, we instantiate our FUDLR framework with the representative LLM-RS, BIGRec \cite{DBLP:journals/tors/BaoZWZYLCFT25}; however, it can be easily adopted to other LLM-RS models, e.g., T5 \cite{DBLP:conf/eacl/HuaGXJLZ24} and TALLRec \cite{Bao2023Tallrec}. Specifically, it first converts the user-item interaction history and the target item $(\mathcal{S}^u_{1:n}, i_{n+1}^u)$ for user $u$ into a natural language prompt $z_k$ (see more details in \cite{DBLP:journals/tors/BaoZWZYLCFT25}) and forms the training set $\mathcal{D}_{\rm train} = \{z_1, z_2, \cdots, z_n\}$. Then, it fine-tunes a pretrained LLM, e.g., LLaMA \cite{DBLP:journals/corr/abs-2407-21783}, on $\mathcal{D}_{\rm train}$ to obtain the recommendation model $\theta$. During the inference stage, we prompt the fine-tuned LLM with the user interaction history and generate the next recommended item $\hat{i}^u_{n+1}=\text{LLM}(z_k; \theta)$. We rank the candidate items based on their distance to the generated output in the embedding space: $d_i = \| \text{Emb}(i) - \text{Emb}(\hat{i}^u_{n+1}) \|,\ \forall i \in \mathcal{I}$, where $\text{Emb}(\cdot)$ denotes the embeddings generated by the LLM.

\textbf{Debiasing LLM-RS.} As LLM-RS are typically trained on large-scale, uncurated datasets, they are prone to inheriting various biases present in the training data. For a pre-trained LLM which is then well fine-tuned on the recommendation dataset $\mathcal{D}_{\rm train}$, classic debiasing methods usually require retraining or fine-tuning the model $\min_{\theta} \mathcal{L}_{\rm LLM}(\mathcal{D}_{\rm train};\theta)$ with specific strategies, such as reweighting \cite{DBLP:conf/www/JiangBZW0F024} the contribution of biased samples to the loss function, so that the constraints on fairness $\mathcal{B}(\theta) \leq \delta$ is satisfied, where $\mathcal{B}(\theta)$ is a fairness metric quantifying the bias level of the trained model $\theta$ and $\delta$ is a predefined threshold. To improve the efficiency without retraining, we reformulate the debiasing LLM-RS task from a machine unlearning perspective. We aim to update the trained model with $\theta + \Delta \theta$ by removing the influence of a small set of bias-inducing training samples $\mathcal{D}_{\rm unlearn} \subset \mathcal{D}_{\rm train}$, such that the updated model shows an improved fairness regarding the metric $\mathcal{B}(\theta + \Delta \theta)$. 

However, this formulation faces two main challenges: 1) How to effectively identify the bias-inducing samples $\mathcal{D}_{\rm unlearn}$ in a bias-agnostic manner? 2) How to efficiently compute the model update $\Delta \theta$ without retraining the LLM-RS from scratch?

\section{The FUDLR Framework}
In this paper, we propose FUDLR, a practical framework for LLM-RS debiasing. It addresses the above challenges through a two-stage process: 1) \textbf{Bias Identification} and 2) \textbf{Debiasing via Unlearning}. First, FUDLR identifies the bias-inducing training samples $\mathcal{D}_{\rm unlearn}$ by learning a novel bias-agnostic mask that optimizes a flexible objective balancing fairness improvement and accuracy preservation. Then, FUDLR performs efficient debiasing by estimating and removing the influence of the identified samples on the model parameters. The overall framework is illustrated in Figure \ref{fig:framework}.

\subsection{Bias Identification}
To identify the bias-inducing training samples from massive training data and maintain high adaptability to various bias types, we design a novel mask learning approach. For each candidate sample $z_k$, it learns a probability $m_k = \sigma(\omega_k) \in [0, 1]$ of the sample being considered for removal, where $\sigma(\cdot)$ is the sigmoid function and $\omega_k$ is a learnable logit. This learnable mask $\mathbf{m}$ is optimized with a flexible objective that jointly balances three goals: 1) maximizing the fairness improvement after unlearning the identified samples; 2) preserving the recommendation accuracy; and 3) ensuring the intervention for mitigating bias is sparse and targeted. 

\textbf{Fairness Improvement.} As the core objective of this stage, we aim to measure and identify the samples that have the most significant influence on the overall model bias. Inspired by the influence function \cite{DBLP:conf/icml/KohL17}, we define an influence score $I(z_k, \mathcal{B}(\theta))$ for each sample $z_k$ to estimate the impact of removing $z_k$ on the fairness metric $\mathcal{B}(\theta)$:
\begin{equation}
\begin{aligned}
I(z_k, \mathcal{B}(\theta)) &= \frac{d\mathcal{B}(\theta_{z_k})}{d\epsilon}\Big|_{\epsilon=0} = \frac{d\mathcal{B}(\theta_{z_k})}{d\theta_{z_k}} \frac{d\theta_{z_k}}{d\epsilon}\Big|_{\epsilon=0} \\
&= -\nabla_{\theta} \mathcal{B}(\theta)^T \mathbf{H}_{\theta}^{-1} \nabla_{\theta} \mathcal{L}_{\text{LLM}}(z_k; \theta),
\label{eq:influence_score}
\end{aligned}
\end{equation}
where $\theta_{z_k}$ represents the new model parameters that would be obtained if the training loss for a single data point $z_k$ were up-weighted by an infinitesimal amount $\epsilon$, and $\mathcal{L}_{\rm LLM}$ is the used fine-tuning loss function. Here, $\mathbf{H}_{\theta} = \frac{1}{n} \sum_{z_k} \nabla_{\theta}^2 \mathcal{L}_{\text{LLM}}(z_k; \theta)$ is the computed Hessian matrix of the empirical loss. A high positive influence score $I(z_k, \mathcal{B}(\theta))$ indicates that removing $z_k$ would lead to a more significant improvement in fairness. Thus, we define the fairness improvement objective as follows:
\begin{equation}
\mathcal{L}_{\rm fair} = - \frac{1}{|\mathcal{D}_{\rm cand}|} \sum_{z_k \in \mathcal{D}_{\rm cand}} m_k \cdot I(z_k, \mathcal{B}(\theta)),
\end{equation}
where $\mathcal{D}_{\rm cand} \subseteq \mathcal{D}_{\rm train}$ is a set of candidate samples for efficiency. In this paper, we use the uniform sampling over the training set to construct $\mathcal{D}_{\rm cand}$ without changing the original data distribution.

\textbf{Accuracy Preservation.} While aiming to improve fairness, it is crucial to preserve reasonable recommendation accuracy of the LLM-RS. Therefore, we introduce an accuracy preservation objective that minimizes the expected training loss over the identified samples to be unlearned:
\begin{equation}
\mathcal{L}_{\rm acc} = \frac{1}{|\mathcal{D}_{\rm cand}|} \sum_{z_k \in \mathcal{D}_{\rm cand}} m_k \cdot \mathcal{L}_{\rm LLM}(z_k; \theta).
\end{equation}
Here, we use the fine-tuning loss $\mathcal{L}_{\rm LLM}$ as a proxy for estimating the contribution of each sample to the overall accuracy. It encourages the mask to avoid selecting samples that are critical for maintaining the balance between fairness and accuracy.

\textbf{Sparsity Regularization.} Sparsity is another key constraint for the mask optimization, as we aim to identify a small and targeted set of bias-inducing samples for efficient unlearning. Meanwhile, existing studies \cite{DBLP:conf/sigir/ZhangF0WSL021} have shown that even biased samples can still contribute positively to the recommendation process. Therefore, to avoid excessive removal of training data and preserve useful information, we introduce a sparse regularization term on the mask:
\begin{equation}
\mathcal{L}_{\rm spa} = \frac{1}{|\mathcal{D}_{\rm cand}|} \sum_{z_k \in \mathcal{D}_{\rm cand}} m_k.
\end{equation}
This term encourages the mask to select only a small fraction of samples for unlearning.

\textbf{Overall Objective.} By combining the above three objectives, we formulate the overall mask learning objective as:
\begin{equation}
\label{eq:Overall_Objective}
\mathcal{L}_{\rm mask} = \lambda_{\rm fair}\mathcal{L}_{\rm fair} + \lambda_{\rm acc} \mathcal{L}_{\rm acc} + \lambda_{\rm spa} \mathcal{L}_{\rm spa},
\end{equation}
where $\lambda_{\rm fair}$, $\lambda_{\rm acc}$ and $\lambda_{\rm spa}$ are hyperparameters that control the trade-off among fairness improvement, accuracy preservation, and sparsity. We optimize the mask logits $\{\omega_k\}$ by minimizing $\mathcal{L}_{\rm mask}$ using gradient descent. After optimization, we select the samples with positive learned logits ($\omega_k > 0$) as the target samples to unlearn: $\mathcal{D}_{\rm unlearn} = \{z_k | \omega_k > 0, z_k \in \mathcal{D}_{\rm cand}\}$.

\subsection{Debiasing via Unlearning}
After identifying the bias-inducing samples $\mathcal{D}_{\rm unlearn}$, an intuitive way to debias the model is to retrain it on the remaining data with the following objective:
\begin{equation}
\scalebox{1}{$
\hat{\theta} = \arg\min_{\theta} \sum_{z_k \in \mathcal{D}_{\rm remain}} \mathcal{L}_{\rm LLM}(z_k; \theta),
$}
\end{equation}
where $\mathcal{D}_{\rm remain} = \mathcal{D}_{\rm train} \setminus \mathcal{D}_{\rm unlearn}$. However, this retraining process is computationally infeasible for large-scale LLMs. In this stage, we aim to compute a parameter update $\Delta \theta$ to estimate the model $\hat{\theta}$ that would have been obtained if these samples were removed from the training set, such that, we have
\begin{equation}
\nabla_\theta \mathcal{L}_{\rm LLM}(\mathcal{D}_{\rm remain}; \theta + \Delta \theta) \approx 0.
\end{equation}
To achieve this, we propose a proposition that leverages influence functions to estimate the parameter update efficiently.
\begin{proposition}
\label{prop:unlearning}
Given a trained biased LLM-RS model parameterized by $\theta$ that minimizes the empirical risk on a training set $\mathcal{D}_{\rm train}$ with $n$ samples, and an identified subset of bias-inducing samples $\mathcal{D}_{\rm unlearn}$, the debiased parameter update $\Delta \theta$ required to approximate the model trained on $\mathcal{D}_{\rm remain} = \mathcal{D}_{\rm train} \setminus \mathcal{D}_{\rm unlearn}$ is given by the aggregated influence of the unlearned data:
\begin{equation}
\scalebox{1}{$
\Delta \theta \approx \frac{1}{n} \sum_{z_k \in \mathcal{D}_{\rm unlearn}} \mathbf{H}_{\theta}^{-1} \nabla_{\theta} \mathcal{L}_{\rm LLM}(z_k; \theta),
$}
\end{equation}
where $\mathbf{H}_{\theta}$ is the invertible Hessian matrix.
\end{proposition}
The proof of Proposition \ref{prop:unlearning} is provided in Appendix \ref{appen:proof}. Based on this proposition, we can efficiently compute the debiased model parameters using 
\begin{equation}
  \theta^* = \theta + \Delta \theta.
\end{equation}
This influence estimation and parameter updates are \textbf{efficient} and \textbf{accurate}, as verified by the negligible performance gap (Appendix \ref{appen:estimation}) between our results and those of the retrained model.

\subsection{Practical Discussion}
\subsubsection{Generalizability to Various Bias Types}
The core of FUDLR's generalizability lies in disentangling the bias identification and mitigation process from specific bias types. By leveraging various differentiable fairness metrics $\mathcal{B}(\theta)$ in the mask learning stage, FUDLR can adapt to various bias types without any other algorithmic changes. Based on the extensive literature on fairness in recommender systems \cite{Wang2023Survey,DBLP:journals/umuai/DeldjooJBDZ24,DBLP:journals/tist/LiCXGTLZ23}, we can easily instantiate FUDLR for different bias types by defining appropriate bias metrics. Here, we take the representative item-side popularity bias and user-side attribute bias as examples.

\textbf{Popularity Bias.} Inspired by the classic popularity bias metric~\cite{DBLP:conf/flairs/AbdollahpouriBM19} that measures the average popularity of recommended items, we can define a differentiable metric for LLM-RS:
\begin{equation}
\scalebox{1}{$
\mathcal{B}_{\rm pop}(z;\theta) = \sum_{i\in\mathcal{I}}P(i|z;\theta) \cdot v_{\rm pop}(i),
$}
\end{equation}
where $v_{\rm pop}(i)$ denotes the popularity value of item $i$ and $P(i|z;\theta)$ is the probability distribution over items given the prompt $z$ generated by the LLM-RS, and is defined as:
\begin{equation}
P(i|z;\theta) = \frac{\exp(-d_i)}{\sum_{j\in\mathcal{I}} \exp(-d_j)}.
\end{equation}

\textbf{Attribute Bias.} We adopt the demographic parity metric \cite{DBLP:journals/umuai/DeldjooJBDZ24} that measures the recommendation rate across different user groups, e.g., the gender groups $G_0$ and $G_1$. By computing the average recommendation probability for each group, $\bar{P}_{G_0}$ and $\bar{P}_{G_1}$, we can define a differentiable attribute bias metric for LLM-RS as:
\begin{equation}
\mathcal{B}_{\rm attr}(\theta) = |\bar{P}_{G_0} - \bar{P}_{G_1}|,
\end{equation}
where $\bar{P}_{G_j} = \frac{1}{|\mathcal{U}_{G_j}|} \sum_{u \in \mathcal{U}_{G_j}} P(i|z_u;\theta)$, $\mathcal{U}_{G_j}$ is the set of users in group $G_j$, and $j=0, 1$.

Moving beyond these examples, FUDLR can be easily extended to other bias types by defining corresponding differentiable metrics, such as equality of opportunity \cite{Chen2023Improving} and exposure fairness \cite{Ge2021Towards}. More importantly, owing to the high flexibility of the mask learning mechanism, FUDLR can also adapt to emerging or \textbf{multiple co-existing biases} by combining multiple bias metrics in the mask learning objective. For instance, to mitigate both popularity and attribute biases simultaneously, we define a combined bias metric:
\begin{equation}
\mathcal{B}_{\rm combined}(\theta) = \alpha \mathcal{B}_{\rm pop}(\theta) + (1-\alpha) \mathcal{B}_{\rm attr}(\theta),
\end{equation}
where $\alpha \in [0, 1]$ controls the trade-off between the two bias types. By substituting $\mathcal{B}(\theta)$ with $\mathcal{B}_{\rm combined}(\theta)$ in the mask learning stage in Eq. \eqref{eq:influence_score}, FUDLR can effectively identify and mitigate samples contributing to both biases.

\subsubsection{Computational Efficiency}
Despite the efficiency of FUDLR compared to retraining-based debiasing methods, directly computing the influence scores and Hessian inverse is still computationally expensive for large-scale LLMs. To further enhance efficiency, we adopt the following practical strategies:

\textbf{Targeting LoRA Adapters.} Instead of applying FUDLR to the full LLM parameters, we focus on the last layer of LoRA adapters \cite{Hu2022Lora} that are fine-tuned for recommendation tasks. LoRA adapters introduce a small number of trainable parameters into the frozen LLM, significantly reducing the parameter space and computational cost for influence estimation.

\textbf{Hessian Matrix Approximation.} To avoid the costly computation of the full Hessian matrix, we approximate it using the Hessian-vector product (HVP) technique \cite{DBLP:conf/icml/KohL17}. This approach allows us to compute the product of the Hessian with a vector without explicitly forming the Hessian matrix, thereby reducing memory and computation requirements. Meanwhile, the influence scores in the bias identification stage can be pre-computed once and reused during mask optimization, further enhancing efficiency.

\subsection{Computational Complexity}
Here, we provide a theoretical analysis of the computational time complexity of FUDLR for each stage.

\textbf{Bias Identification.} The main computational cost in this stage arises from calculating the influence scores for the candidate samples. Let $n_c=|\mathcal{D}_{\rm cand}|$, $n_u=|\mathcal{D}_{\rm unlearn}|$, and $F$ denotes the cost of the LLM forward pass and gradient computation for the adapter parameters. The time complexity for computing $\nabla_\theta\mathcal{B}(\theta)$ is $O(n_c \cdot F)$. Solving for $\mathbf{H}_\theta^{-1} \nabla_\theta \mathcal{B}(\theta)$ using HVP typically requires $O(T_{\rm cg} \cdot F)$ with $T_{\rm cg}$ iterations. Putting these together, the dominant time complexity for bias identification is $O((n_c + T_{\rm cg}) \cdot F)$.

\textbf{Debiasing via Unlearning.} In this stage, the main cost comes from computing the aggregated influence of the unlearned samples. Similar to the previous stage, computing $\nabla_\theta \mathcal{L}_{\rm LLM}(z_k; \theta)$ for each unlearned sample requires $O(F)$ time. Thus, the total time complexity for debiasing via unlearning is $O((n_u + T_{\rm cg}) \cdot F)$.

Since the identified unlearning set is typically a small subset of the training data due to the sparsity constraint, thus, we have $n_u \ll n$, and the overall time complexity of FUDLR is significantly lower than retraining-based methods with a complexity of $O(n \cdot E \cdot F)$, where $E$ is the number of training epochs. Therefore, FUDLR offers a computationally efficient solution for debiasing LLM-RS.

\section{Experiments}
In this section, we design a series of experiments to explore the following research questions (RQs):
\begin{itemize}[left=0pt]
  \item \textbf{RQ1}: How does FUDLR perform in mitigating representative item-side bias, e.g., popularity bias, in LLM-RS compared to state-of-the-art baselines?
  \item \textbf{RQ2}: How does FUDLR perform in mitigating representative user-side bias, e.g., attribute bias, in LLM-RS compared to state-of-the-art baselines?
  \item \textbf{RQ3}: How does FUDLR perform in mitigating multiple co-existing types of bias in LLM-RS?
  \item \textbf{RQ4}: How do the key components and hyperparameters of FUDLR affect its performance?
\end{itemize}

\subsection{Experimental Setup}
\subsubsection{Datasets}
We adopt two real-world datasets widely-used in LLM-RS to evaluate the effectiveness of debiasing methods. 1) \textbf{MovieLens1M}\footnote{\url{https://grouplens.org/datasets/movielens/1m/}} (\textbf{ML1M}) \cite{DBLP:conf/www/JiangBZW0F024} is a popular benchmark dataset for movie recommendation. It is also widely used for studying popularity bias \cite{DBLP:conf/www/JiangBZW0F024,Deldjoo2024Understanding,DBLP:conf/flairs/AbdollahpouriBM19} and attribute bias \cite{Deldjoo2024Understanding,Chen2023Improving} in RS for its long-tailed item popularity distribution \cite{Abdollahpouri2017Controlling} and rich user demographic information. 2) \textbf{Games}\footnote{\url{https://jmcauley.ucsd.edu/data/amazon/}} \cite{DBLP:journals/tors/BaoZWZYLCFT25,DBLP:conf/www/JiangBZW0F024} is a subset of the Amazon Review dataset \cite{Ni2019Justifying} that contains ratings for video games. The statistics of the datasets used in our experiments are summarized in Appendix \ref{appen:Datasets details}.

We follow the preprocessing steps in \cite{DBLP:journals/tors/BaoZWZYLCFT25} for both datasets. Specifically, we divide each dataset into 10 periods based on the timestamp of the interactions to prevent data leakage. We further split each dataset into training, validation, and test sets with a ratio of 8:1:1. Following the common practice in LLM-RS \cite{DBLP:journals/tors/BaoZWZYLCFT25,DBLP:conf/www/JiangBZW0F024}, we sample 65,536 instances for training and 5,000 instances for testing without altering the distribution of the original training dataset.

\subsubsection{Baselines}
To evaluate the debiasing effectiveness on representative biases, e.g., item-side popularity bias (\textbf{RQ1}) and user-side attribute bias (\textbf{RQ2}), we compare FUDLR with the following state-of-the-art baselines regarding their performance and time costs, including a debiasing time (e.g., retraining in baselines or bias identification and mitigation in FUDLR) and an inference time.

1) \textbf{Popularity Debiasing Baselines:} We include three representative strategies: (a) \textbf{Reweighting} \cite{DBLP:conf/www/JiangBZW0F024} adjusts the loss contribution of each training sample based on item popularity during retraining; (b) \textbf{Reranking} \cite{DBLP:conf/www/JiangBZW0F024} applies a post-processing step to rerank recommended items by introducing a penalty term for popular items; (c) \textbf{Reweighting + Reranking (RWRR)} \cite{DBLP:conf/www/JiangBZW0F024} combines both reweighting and reranking strategies for enhanced debiasing. 

2) \textbf{Attribute Debiasing Baselines:} We consider two representative methods: (a) \textbf{Counterfactually-Fair-Prompt (CFP)} \cite{DBLP:conf/eacl/HuaGXJLZ24} employs adversarial learning to eliminate sensitive information from token embeddings during fine-tuning; (b) \textbf{Prompt Masking (Masking)} \cite{DBLP:journals/ipm/ShenLBMS23} removes sensitive attributes from input prompts.

Given that research on LLM-RS debiasing is still in its early stages and conventional RS debiasing methods \cite{DBLP:journals/tois/ShaoWZLHLW24,DBLP:conf/sigir/ZhangF0WSL021,DBLP:conf/www/ZhengGLHLJ21,DBLP:conf/sigir/ChenDQ0XCLY21} often require specific adaptations for LLM-RS, the available baseline methods are limited. Nevertheless, we have carefully selected representative baselines to ensure a comprehensive evaluation.

\subsubsection{Evaluation Metrics}
To comprehensively evaluate the performance of FUDLR and baselines regarding both recommendation accuracy and fairness, we adopt the following metrics: 1) \textbf{Accuracy Metrics:} We use Hit Rate (\textbf{HR@K}) and Normalized Discounted Cumulative Gain (\textbf{NDCG@K}) to measure the recommendation accuracy. The higher the values, the better the accuracy. 2) \textbf{Fairness Metrics:} For popularity bias, we use Average Recommendation Popularity (\textbf{ARP}) \cite{DBLP:conf/flairs/AbdollahpouriBM19,Abdollahpouri2017Controlling} and Average Percentage of Long Tail Items (\textbf{APT}) \cite{DBLP:conf/flairs/AbdollahpouriBM19,Abdollahpouri2017Controlling} to quantify the average popularity and long-tail item exposure in recommendations, respectively. The higher the values, the better the fairness. For attribute bias, we use Hit Rate Difference (\textbf{HD}) and Demographic Parity (\textbf{DP}) \cite{wang-etal-2023-improving-conversational} to measure the fairness of recommendations towards different user groups. For both metrics, the lower the values, the better the fairness. 3) \textbf{Combined Metric:} To evaluate the overall performance considering both accuracy and fairness, we adopt the F1 score that combines HR@10 and ARP for popularity bias, and HR@10 and HD for attribute bias.
\begin{equation}
F_{\rm pop}@K = 2 \cdot \frac{\tau{\rm HR@K} \cdot {\rm Fair}@K}{\tau{\rm HR@K} + {\rm Fair}@K},
\end{equation}
where $\tau$ is a balancing parameter set to 5 to scale HR@K to a similar range as Fair@K. For popularity bias, we use ${\rm Fair}@K = {\rm ARP}@K$; for attribute bias, we use ${\rm Fair}@K = 1 - {\rm HD}@K$.

\subsubsection{Implementation Details}
We implement FUDLR based on the representative BIGRec \cite{DBLP:journals/tors/BaoZWZYLCFT25} framework using LLaMA 3.1 8B as the backbone. During mask learning, we sample the candidate set with a uniform distribution over the training set with a ratio of 10\% without changing the original data distribution. The hyperparameters $\lambda_{\rm fair}$, $\lambda_{\rm acc}$, and $\lambda_{\rm spa}$ are tuned via grid search within $\{10^i\}_{i=-3}^{2}$. We use Adam optimizer with a learning rate of $10^{-3}$ for mask optimization. All experiments are conducted on clusters with 2 Intel Xeon Gold 6346 CPUs, 256GB RAM, and 2 NVIDIA A40 GPUs.

\subsection{RQ1: Popularity Bias Mitigation}
We evaluate the performance of FUDLR in popularity debiasing in the ML1M and Games datasets. Following the common practice in popularity debiasing research \cite{Abdollahpouri2017Controlling}, we classify items into two groups based on their popularity: the short head (the top 20\% most popular items) and the long tail (the remaining 80\% items). Results are shown in Table \ref{tab:pop}, from which we have the following observations.

First, compared to the backbone model (BIGRec), the proposed FUDLR consistently improves both fairness and accuracy. This highlights FUDLR's effectiveness in mitigating popularity bias while successfully balancing these two key objectives. 

Second, when compared to state-of-the-art popularity debiasing methods, FUDLR achieves superior performance in terms of both accuracy and fairness in most cases. Although debiasing baselines achieve higher ARP and APT scores in the ML1M dataset, they suffer from severe accuracy degradation. As shown in the case study (Appendix \ref{appen:case_study}), methods like RWRR, while recommending debiased results, often diverge substantially from users’ genuine preferences, which should be the core objective of RS. In contrast, our method delivers not only debiased but also accurate results.

Third, rather than requiring costly retraining, FUDLR directly estimates the influence of bias-inducing samples and updates parameters for debiasing. This significantly reduces computational costs. For instance, FUDLR reduces runtime by 96.27\% compared to the backbone and is approximately 32 times faster than RWRR in the ML1M dataset. Although the Reranking method is slightly faster than FUDLR (about 0.17 hours less), it is a post-processing strategy that usually fails to optimally balance accuracy and fairness.

Overall, FUDLR has shown effectiveness in striking an optimal balance between fairness enhancement and accuracy preservation with remarkable efficiency.
\vspace{-5pt}

\begin{table*}[ht]
  \centering
  \caption{Performance comparison for popularity bias mitigation on ML1M and Games datasets. The best results are in \textbf{bold}. The Improve is calculated as the relative improvement of each method over BIGRec. The \textcolor[rgb]{1,0,0}{red} Improve values indicate improved performance, while the \textcolor[rgb]{ 0,  .439,  .753}{blue} ones indicate degraded performance. *the improvement is significant at p<0.05.}
  \vspace{-5pt}
  \renewcommand{\arraystretch}{0.95}
  \resizebox{\linewidth}{!}{
    \begin{tabular}{c|c|cc|cc|cc|cc|cc|c}
    \hline
    \multirow{2}[2]{*}{Datasets} & \multirow{2}[2]{*}{Methods} & \multicolumn{2}{c|}{HR$\uparrow$} & \multicolumn{2}{c|}{NDCG$\uparrow$} & \multicolumn{2}{c|}{ARP$\uparrow$} & \multicolumn{2}{c|}{APT$\uparrow$} & \multicolumn{2}{c|}{$F_{\rm pop}$$\uparrow$} & \multirow{2}[2]{*}{Time (h) $\downarrow$} \\
          &       & K=5   & K=20  & K=5   & K=20  & K=5   & K=20  & K=5   & K=20  & K=5   & K=20  &  \\
    \hline
    \multirow{9}[2]{*}{ML1M} & BIGRec & 0.0220 & 0.0336 & 0.0167 & 0.0200 & 5.0237 & 4.7246 & 0.3256 & 0.2476 & 0.1644 & 0.2002 & 34.79 \\ 
    \cline{2-13}          & Reweighting & 0.0180 & 0.0296 & 0.0131 & 0.0163 & \textbf{5.4806} & 4.9093 & 0.4529 & 0.2883 & 0.1502 & 0.1956 & 40.95 \\
          & \cellcolor[rgb]{ .851,  .851,  .851}\textit{Improve (\%)} & \cellcolor[rgb]{ .851,  .851,  .851}\textcolor[rgb]{ 0,  .439,  .753}{-18.18} & \cellcolor[rgb]{ .851,  .851,  .851}\textcolor[rgb]{ 0,  .439,  .753}{-11.90} & \cellcolor[rgb]{ .851,  .851,  .851}\textcolor[rgb]{ 0,  .439,  .753}{-21.69} & \cellcolor[rgb]{ .851,  .851,  .851}\textcolor[rgb]{ 0,  .439,  .753}{-18.32} & \cellcolor[rgb]{ .851,  .851,  .851}\textcolor[rgb]{ 1,  0,  0}{9.09} & \cellcolor[rgb]{ .851,  .851,  .851}\textcolor[rgb]{ 1,  0,  0}{3.91} & \cellcolor[rgb]{ .851,  .851,  .851}\textcolor[rgb]{ 1,  0,  0}{39.11} & \cellcolor[rgb]{ .851,  .851,  .851}\textcolor[rgb]{ 1,  0,  0}{16.42} & \cellcolor[rgb]{ .851,  .851,  .851}\textcolor[rgb]{ 0,  .439,  .753}{-8.68} & \cellcolor[rgb]{ .851,  .851,  .851}\textcolor[rgb]{ 0,  .439,  .753}{-2.30} & \cellcolor[rgb]{ .851,  .851,  .851} \textcolor[rgb]{ 0,  .439,  .753}{-17.70} \\
          & Reranking & 0.0102 & 0.0230 & 0.0080 & 0.0114 & 4.8443 & 4.8577 & 0.4082 & 0.3159 & 0.0907 & 0.1686 & \textbf{1.12} \\
          & \cellcolor[rgb]{ .851,  .851,  .851}\textit{Improve} & \cellcolor[rgb]{ .851,  .851,  .851}\textcolor[rgb]{ 0,  .439,  .753}{-53.64} & \cellcolor[rgb]{ .851,  .851,  .851}\textcolor[rgb]{ 0,  .439,  .753}{-31.55} & \cellcolor[rgb]{ .851,  .851,  .851}\textcolor[rgb]{ 0,  .439,  .753}{-52.18} & \cellcolor[rgb]{ .851,  .851,  .851}\textcolor[rgb]{ 0,  .439,  .753}{-42.88} & \cellcolor[rgb]{ .851,  .851,  .851}\textcolor[rgb]{ 0,  .439,  .753}{-3.57} & \cellcolor[rgb]{ .851,  .851,  .851}\textcolor[rgb]{ 1,  0,  0}{2.82} & \cellcolor[rgb]{ .851,  .851,  .851}\textcolor[rgb]{ 1,  0,  0}{25.38} & \cellcolor[rgb]{ .851,  .851,  .851}\textcolor[rgb]{ 1,  0,  0}{27.56} & \cellcolor[rgb]{ .851,  .851,  .851}\textcolor[rgb]{ 0,  .439,  .753}{-44.86} & \cellcolor[rgb]{ .851,  .851,  .851}\textcolor[rgb]{ 0,  .439,  .753}{-15.77} & \cellcolor[rgb]{ .851,  .851,  .851} \textcolor[rgb]{ 1,  0,  0}{96.78}  \\
          & RWRR  & 0.0084 & 0.0214 & 0.0060 & 0.0096 & 5.2571 & \textbf{5.0690} & \textbf{0.5013} & \textbf{0.3577} & 0.0775 & 0.1647 & 41.6 \\
          & \cellcolor[rgb]{ .851,  .851,  .851}\textit{Improve (\%)} & \cellcolor[rgb]{ .851,  .851,  .851}\textcolor[rgb]{ 0,  .439,  .753}{-61.82} & \cellcolor[rgb]{ .851,  .851,  .851}\textcolor[rgb]{ 0,  .439,  .753}{-36.31} & \cellcolor[rgb]{ .851,  .851,  .851}\textcolor[rgb]{ 0,  .439,  .753}{-64.13} & \cellcolor[rgb]{ .851,  .851,  .851}\textcolor[rgb]{ 0,  .439,  .753}{-51.90} & \cellcolor[rgb]{ .851,  .851,  .851}\textcolor[rgb]{ 1,  0,  0}{4.65} & \cellcolor[rgb]{ .851,  .851,  .851}\textcolor[rgb]{ 1,  0,  0}{7.29} & \cellcolor[rgb]{ .851,  .851,  .851}\textcolor[rgb]{ 1,  0,  0}{53.98} & \cellcolor[rgb]{ .851,  .851,  .851}\textcolor[rgb]{ 1,  0,  0}{44.44} & \cellcolor[rgb]{ .851,  .851,  .851}\textcolor[rgb]{ 0,  .439,  .753}{-52.87} & \cellcolor[rgb]{ .851,  .851,  .851}\textcolor[rgb]{ 0,  .439,  .753}{-17.72} & \cellcolor[rgb]{ .851,  .851,  .851} \textcolor[rgb]{ 0,  .439,  .753}{-19.57}\\
          & FUDLR  & \textbf{0.0226*} & \textbf{0.0340*} & \textbf{0.0172*} & \textbf{0.0203*} & 5.0310 & 4.7276 & 0.3278 & 0.2486 & \textbf{0.1681*} & \textbf{0.2019*} & 1.29 \\
          & \cellcolor[rgb]{ .851,  .851,  .851}\textit{Improve (\%)} & \cellcolor[rgb]{ .851,  .851,  .851}\textcolor[rgb]{ 1,  0,  0}{2.65} & \cellcolor[rgb]{ .851,  .851,  .851}\textcolor[rgb]{ 1,  0,  0}{1.18} & \cellcolor[rgb]{ .851,  .851,  .851}\textcolor[rgb]{ 1,  0,  0}{2.63} & \cellcolor[rgb]{ .851,  .851,  .851}\textcolor[rgb]{ 1,  0,  0}{1.91} & \cellcolor[rgb]{ .851,  .851,  .851}\textcolor[rgb]{ 1,  0,  0}{0.14} & \cellcolor[rgb]{ .851,  .851,  .851}\textcolor[rgb]{ 1,  0,  0}{0.06} & \cellcolor[rgb]{ .851,  .851,  .851}\textcolor[rgb]{ 1,  0,  0}{0.67} & \cellcolor[rgb]{ .851,  .851,  .851}\textcolor[rgb]{ 1,  0,  0}{0.37} & \cellcolor[rgb]{ .851,  .851,  .851}\textcolor[rgb]{ 1,  0,  0}{2.15} & \cellcolor[rgb]{ .851,  .851,  .851}\textcolor[rgb]{ 1,  0,  0}{0.85} & \cellcolor[rgb]{ .851,  .851,  .851}\textcolor[rgb]{1,0,0}{96.27}\\ 
    \hline
    \multirow{9}[2]{*}{Games} & BIGRec & 0.0304 & 0.0488 & 0.0244 & 0.0295 & 3.6487 & 3.3519 & 0.4159 & 0.3242 & 0.2226 & 0.2784 & 41.08 \\
    \cline{2-13} 
          & Reweighting & 0.0135 & 0.0270 & 0.0102 & 0.0139 & 3.4551 & 3.2980 & 0.3526 & 0.2995 & 0.1133 & 0.1861 & 48.36 \\
          & \cellcolor[rgb]{ .851,  .851,  .851}\textit{Improve (\%)} & \cellcolor[rgb]{ .851,  .851,  .851}\textcolor[rgb]{ 0,  .439,  .753}{-55.59} & \cellcolor[rgb]{ .851,  .851,  .851}\textcolor[rgb]{ 0,  .439,  .753}{-44.67} & \cellcolor[rgb]{ .851,  .851,  .851}\textcolor[rgb]{ 0,  .439,  .753}{-58.16} & \cellcolor[rgb]{ .851,  .851,  .851}\textcolor[rgb]{ 0,  .439,  .753}{-52.91} & \cellcolor[rgb]{ .851,  .851,  .851}\textcolor[rgb]{ 0,  .439,  .753}{-5.31} & \cellcolor[rgb]{ .851,  .851,  .851}\textcolor[rgb]{ 0,  .439,  .753}{-1.61} & \cellcolor[rgb]{ .851,  .851,  .851}\textcolor[rgb]{ 0,  .439,  .753}{-15.22} & \cellcolor[rgb]{ .851,  .851,  .851}\textcolor[rgb]{ 0,  .439,  .753}{-7.61} & \cellcolor[rgb]{ .851,  .851,  .851}\textcolor[rgb]{ 0,  .439,  .753}{-49.11} & \cellcolor[rgb]{ .851,  .851,  .851}\textcolor[rgb]{ 0,  .439,  .753}{-33.16} & \cellcolor[rgb]{ .851,  .851,  .851} \textcolor[rgb]{ 0,  .439,  .753}{-17.72}\\
          & Reranking & 0.0156 & 0.0250 & 0.0118 & 0.0144 & 3.4572 & 3.0548 & 0.3923 & 0.2550 & 0.1301 & 0.1678 & \textbf{1.32} \\
          & \cellcolor[rgb]{ .851,  .851,  .851}\textit{Improve (\%)} & \cellcolor[rgb]{ .851,  .851,  .851}\textcolor[rgb]{ 0,  .439,  .753}{-48.68} & \cellcolor[rgb]{ .851,  .851,  .851}\textcolor[rgb]{ 0,  .439,  .753}{-48.77} & \cellcolor[rgb]{ .851,  .851,  .851}\textcolor[rgb]{ 0,  .439,  .753}{-51.60} & \cellcolor[rgb]{ .851,  .851,  .851}\textcolor[rgb]{ 0,  .439,  .753}{-51.21} & \cellcolor[rgb]{ .851,  .851,  .851}\textcolor[rgb]{ 0,  .439,  .753}{-5.25} & \cellcolor[rgb]{ .851,  .851,  .851}\textcolor[rgb]{ 0,  .439,  .753}{-8.86} & \cellcolor[rgb]{ .851,  .851,  .851}\textcolor[rgb]{ 0,  .439,  .753}{-5.68} & \cellcolor[rgb]{ .851,  .851,  .851}\textcolor[rgb]{ 0,  .439,  .753}{-21.34} & \cellcolor[rgb]{ .851,  .851,  .851}\textcolor[rgb]{ 0,  .439,  .753}{-41.55} & \cellcolor[rgb]{ .851,  .851,  .851}\textcolor[rgb]{ 0,  .439,  .753}{-39.75} & \cellcolor[rgb]{ .851,  .851,  .851} \textcolor[rgb]{ 1,  0,  0}{96.79}\\
          & RWRR  & 0.0126 & 0.0236 & 0.0108 & 0.0140 & 3.5179 & 3.1006 & 0.4083 & 0.2660 & 0.1092 & 0.1635 & 49.13 \\
          & \cellcolor[rgb]{ .851,  .851,  .851}\textit{Improve (\%)} & \cellcolor[rgb]{ .851,  .851,  .851}\textcolor[rgb]{ 0,  .439,  .753}{-58.55} & \cellcolor[rgb]{ .851,  .851,  .851}\textcolor[rgb]{ 0,  .439,  .753}{-51.64} & \cellcolor[rgb]{ .851,  .851,  .851}\textcolor[rgb]{ 0,  .439,  .753}{-55.70} & \cellcolor[rgb]{ .851,  .851,  .851}\textcolor[rgb]{ 0,  .439,  .753}{-52.57} & \cellcolor[rgb]{ .851,  .851,  .851}\textcolor[rgb]{ 0,  .439,  .753}{-3.59} & \cellcolor[rgb]{ .851,  .851,  .851}\textcolor[rgb]{ 0,  .439,  .753}{-7.50} & \cellcolor[rgb]{ .851,  .851,  .851}\textcolor[rgb]{ 0,  .439,  .753}{-1.83} & \cellcolor[rgb]{ .851,  .851,  .851}\textcolor[rgb]{ 0,  .439,  .753}{-17.94} & \cellcolor[rgb]{ .851,  .851,  .851}\textcolor[rgb]{ 0,  .439,  .753}{-50.97} & \cellcolor[rgb]{ .851,  .851,  .851}\textcolor[rgb]{ 0,  .439,  .753}{-41.29} & \cellcolor[rgb]{ .851,  .851,  .851}\textcolor[rgb]{ 0,  .439,  .753}{-19.60}\\
          & FUDLR  & \textbf{0.0312*} & \textbf{0.0500*} & \textbf{0.0249*} & \textbf{0.0301*} & \textbf{3.6688*} & \textbf{3.3589*} & \textbf{0.4206*} & \textbf{0.3254*} & \textbf{0.2276*} & \textbf{0.2828*} & 1.49 \\
          & \cellcolor[rgb]{ .851,  .851,  .851}\textit{Improve (\%)} & \cellcolor[rgb]{ .851,  .851,  .851}\textcolor[rgb]{ 1,  0,  0}{2.63} & \cellcolor[rgb]{ .851,  .851,  .851}\textcolor[rgb]{ 1,  0,  0}{2.46} & \cellcolor[rgb]{ .851,  .851,  .851}\textcolor[rgb]{ 1,  0,  0}{2.31} & \cellcolor[rgb]{ .851,  .851,  .851}\textcolor[rgb]{ 1,  0,  0}{2.14} & \cellcolor[rgb]{ .851,  .851,  .851}\textcolor[rgb]{ 1,  0,  0}{0.55} & \cellcolor[rgb]{ .851,  .851,  .851}\textcolor[rgb]{ 1,  0,  0}{0.21} & \cellcolor[rgb]{ .851,  .851,  .851}\textcolor[rgb]{ 1,  0,  0}{1.12} & \cellcolor[rgb]{ .851,  .851,  .851}\textcolor[rgb]{ 1,  0,  0}{0.38} & \cellcolor[rgb]{ .851,  .851,  .851}\textcolor[rgb]{ 1,  0,  0}{2.22} & \cellcolor[rgb]{ .851,  .851,  .851}\textcolor[rgb]{ 1,  0,  0}{1.56} & \cellcolor[rgb]{ .851,  .851,  .851}\textcolor[rgb]{ 1,0,0}{96.36} \\
    \hline
    \end{tabular}}
    \vspace{-5pt}
  \label{tab:pop}
\end{table*}

\subsection{RQ2: Attribute Bias Mitigation}
To assess the generalizability of our framework, we evaluate its performance in alleviating user-side attribute bias (based on gender) in the ML1M dataset. Following widely adopted settings, we evaluate debiasing performance in two scenarios: implicit \cite{DBLP:conf/eacl/HuaGXJLZ24,DBLP:conf/sigir/LiCXGZ21}, where user attributes are not exposed in the prompt, and explicit \cite{DBLP:journals/ipm/ShenLBMS23,DBLP:conf/recsys/ZhangBZWF023,DBLP:conf/recsys/Tommasel24}, where they are. The results are detailed in Table \ref{tab:attr}.

Consistent with observations regarding popularity bias, accuracy-oriented models (such as BIGRec) exhibit unfairness, evidenced by relatively high HD and DP scores, despite achieving strong accuracy. This confirms the presence of attribute bias in the datasets and the models’ tendency to perpetuate it. Among the debiasing baselines, CFP demonstrates inferior performance regarding both recommendation accuracy and fairness metrics in the implicit scenario (e.g., changing by 33.93\% compared to the base model’s HD@20).

In contrast, FUDLR achieves the best results in the vast majority of cases across both scenarios, with the exception of HD@20 in the explicit context. In the implicit scenario, FUDLR not only improves fairness (achieving the lowest HD and DP) but also enhances recommendation accuracy (achieving the highest HR, NDCG) compared to all other methods, including accuracy-oriented baselines. This suggests that the objective of reducing unfairness between user groups aligns with improving overall recommendation quality on this dataset. In the explicit scenario, FUDLR again achieves the optimal balance, recording the highest {$F_{\rm attr}$ scores.

Regarding computational cost, FUDLR achieves dramatically lower runtime than most baselines (e.g., decreased by 95.57\% in the implicit scenario). Although the Masking method is slightly faster in the explicit scenario, its performance is significantly inferior to FUDLR. This underscores FUDLR as an efficient and practical framework to address attribute bias while achieving high accuracy.

\begin{table*}[t]
  \centering
  \caption{Performance comparison for attribute debiasing in implicit and explicit scenarios in the ML1M dataset. The best results are in \textbf{bold}. The Improve is the relative improvement of each method over BIGRec. The \textcolor[rgb]{1,0,0}{red} Improve values indicate improved performance, while the \textcolor[rgb]{ 0,  .439,  .753}{blue} ones indicate degraded performance. *the improvement is significant at p<0.05.}
  \renewcommand{\arraystretch}{0.92}
  \vspace{-5pt}
  \resizebox{\linewidth}{!}{
    \begin{tabular}{c|c|cc|cc|cc|cc|cc|c}
    \hline
    \multirow{2}[2]{*}{Scenarios} & \multirow{2}[2]{*}{Methods} & \multicolumn{2}{c|}{HR$\uparrow$} & \multicolumn{2}{c|}{NDCG$\uparrow$} & \multicolumn{2}{c|}{HD$\downarrow$} & \multicolumn{2}{c|}{DP$\downarrow$} & \multicolumn{2}{c|}{$F_{\rm attr}\uparrow$} & \multirow{2}[2]{*}{Time (h) $\downarrow$} \\
          &       & K=5   & K=20  & K=5   & K=20  & K=5   & K=20  & K=5   & K=20  & K=5   & K=20  &  \\
    \hline
    \multirow{5}[2]{*}{Implicit} & BIGRec & 0.0220 & 0.0336 & 0.0167 & 0.0200 & 0.0083 & 0.0093 & 0.1551 & 0.0974 & 0.1947 & 0.2833 & 34.79 \\
    \cline{2-13} 
          & CFP   & 0.0198 & 0.0316 & 0.0151 & 0.0184 & 0.0092 & 0.0124 & 0.1813 & 0.1198 & 0.1766 & 0.2679 & 58.88 \\
          & \cellcolor[rgb]{ .851,  .851,  .851}\textit{Improve (\%)} & \cellcolor[rgb]{ .851,  .851,  .851}\textcolor[rgb]{ 0,  .439,  .753}{-10.00} & \cellcolor[rgb]{ .851,  .851,  .851}\textcolor[rgb]{ 0,  .439,  .753}{-5.95} & \cellcolor[rgb]{ .851,  .851,  .851}\textcolor[rgb]{ 0,  .439,  .753}{-9.74} & \cellcolor[rgb]{ .851,  .851,  .851}\textcolor[rgb]{ 0,  .439,  .753}{-7.80} & \cellcolor[rgb]{ .851,  .851,  .851}\textcolor[rgb]{ 0,  .439,  .753}{-10.92} & \cellcolor[rgb]{ .851,  .851,  .851}\textcolor[rgb]{ 0,  .439,  .753}{-33.93} & \cellcolor[rgb]{ .851,  .851,  .851}\textcolor[rgb]{ 0,  .439,  .753}{-16.90} & \cellcolor[rgb]{ .851,  .851,  .851}\textcolor[rgb]{ 0,  .439,  .753}{-23.05} & \cellcolor[rgb]{ .851,  .851,  .851}\textcolor[rgb]{ 0,  .439,  .753}{-9.26} & \cellcolor[rgb]{ .851,  .851,  .851}\textcolor[rgb]{ 0,  .439,  .753}{-5.42} & \cellcolor[rgb]{ .851,  .851,  .851}\textcolor[rgb]{ 0,  .439,  .753}{-69.24} \\
          & FUDLR  & \textbf{0.0226*} & \textbf{0.0340*} & \textbf{0.0175*} & \textbf{0.0206*} & \textbf{0.0081*} & \textbf{0.0058*} & \textbf{0.1550*} & \textbf{0.0961*} & \textbf{0.1993*} & \textbf{0.2862*} & \textbf{1.54} \\
          & \cellcolor[rgb]{ .851,  .851,  .851}\textit{Improve (\%)} & \cellcolor[rgb]{ .851,  .851,  .851}\textcolor[rgb]{ 1,  0,  0}{2.73} & \cellcolor[rgb]{ .851,  .851,  .851}\textcolor[rgb]{ 1,  0,  0}{1.19} & \cellcolor[rgb]{ .851,  .851,  .851}\textcolor[rgb]{ 1,  0,  0}{4.63} & \cellcolor[rgb]{ .851,  .851,  .851}\textcolor[rgb]{ 1,  0,  0}{3.29} & \cellcolor[rgb]{ .851,  .851,  .851}\textcolor[rgb]{ 1,  0,  0}{2.11} & \cellcolor[rgb]{ .851,  .851,  .851}\textcolor[rgb]{ 1,  0,  0}{37.36} & \cellcolor[rgb]{ .851,  .851,  .851}\textcolor[rgb]{ 1,  0,  0}{0.08} & \cellcolor[rgb]{ .851,  .851,  .851}\textcolor[rgb]{ 1,  0,  0}{1.32} & \cellcolor[rgb]{ .851,  .851,  .851}\textcolor[rgb]{ 1,  0,  0}{2.41} & \cellcolor[rgb]{ .851,  .851,  .851}\textcolor[rgb]{ 1,  0,  0}{1.02} & \cellcolor[rgb]{ .851,  .851,  .851}\textcolor[rgb]{ 1, 0,  0}{95.57} \\
    \hline
    \multirow{5}[2]{*}{Explicit} & BIGRec & 0.0198 & 0.0332 & 0.0161 & 0.0198 & 0.0033 & \textbf{0.0007} & 0.1554 & 0.0942 & 0.1772 & 0.2806 & 35.48 \\
    \cline{2-13} 
          & Masking  & 0.0176 & 0.0292 & 0.0134 & 0.0166 & 0.0012 & 0.0011 & 0.1573 & 0.0974 & 0.1594 & 0.2513 & \textbf{0.69} \\
          & \cellcolor[rgb]{ .851,  .851,  .851}\textit{Improve (\%)} & \cellcolor[rgb]{ .851,  .851,  .851}\textcolor[rgb]{ 0,  .439,  .753}{-11.11} & \cellcolor[rgb]{ .851,  .851,  .851}\textcolor[rgb]{ 0,  .439,  .753}{-12.05} & \cellcolor[rgb]{ .851,  .851,  .851}\textcolor[rgb]{ 0,  .439,  .753}{-16.78} & \cellcolor[rgb]{ .851,  .851,  .851}\textcolor[rgb]{ 0,  .439,  .753}{-16.19} & \cellcolor[rgb]{ .851,  .851,  .851}\textcolor[rgb]{ 1,  0,  0}{63.12} & \cellcolor[rgb]{ .851,  .851,  .851}\textcolor[rgb]{ 0,  .439,  .753}{-60.58} & \cellcolor[rgb]{ .851,  .851,  .851}\textcolor[rgb]{ 0,  .439,  .753}{-1.22} & \cellcolor[rgb]{ .851,  .851,  .851}\textcolor[rgb]{ 0,  .439,  .753}{-3.38} & \cellcolor[rgb]{ .851,  .851,  .851}\textcolor[rgb]{ 0,  .439,  .753}{-10.08} & \cellcolor[rgb]{ .851,  .851,  .851}\textcolor[rgb]{ 0,  .439,  .753}{-10.42} & \cellcolor[rgb]{ .851,  .851,  .851}\textcolor[rgb]{1,  0,  0}{98.06} \\
          & FUDLR  & \textbf{0.0204*} & \textbf{0.0342*} & \textbf{0.0162*} & \textbf{0.0200*} & \textbf{0.0001*} & 0.0050 & \textbf{0.1468*} & \textbf{0.0918*} & \textbf{0.1822*} & \textbf{0.2878*} & 1.40 \\
          & \cellcolor[rgb]{ .851,  .851,  .851}\textit{Improve(\%)} & \cellcolor[rgb]{ .851,  .851,  .851}\textcolor[rgb]{ 1,  0,  0}{3.03} & \cellcolor[rgb]{ .851,  .851,  .851}\textcolor[rgb]{ 1,  0,  0}{3.01} & \cellcolor[rgb]{ .851,  .851,  .851}\textcolor[rgb]{ 1,  0,  0}{0.51} & \cellcolor[rgb]{ .851,  .851,  .851}\textcolor[rgb]{ 1,  0,  0}{1.02} & \cellcolor[rgb]{ .851,  .851,  .851}\textcolor[rgb]{ 1,  0,  0}{97.83} & \cellcolor[rgb]{ .851,  .851,  .851}\textcolor[rgb]{ 0,  .439,  .753}{-623.19} & \cellcolor[rgb]{ .851,  .851,  .851}\textcolor[rgb]{ 1,  0,  0}{5.55} & \cellcolor[rgb]{ .851,  .851,  .851}\textcolor[rgb]{ 1,  0,  0}{2.57} & \cellcolor[rgb]{ .851,  .851,  .851}\textcolor[rgb]{ 1,  0,  0}{2.82} & \cellcolor[rgb]{ .851,  .851,  .851}\textcolor[rgb]{ 1,  0,  0}{2.58} & \cellcolor[rgb]{ .851,  .851,  .851}\textcolor[rgb]{1,0,0}{96.05} \\
    \hline 
    \end{tabular}}
    \vspace{-5pt}
  \label{tab:attr}
\end{table*}

\subsection{RQ3: Multifaceted Bias Mitigation}
Apart from single bias problems, LLM-RS is usually involved with complex, emerging, and co-existing biases. We evaluate the debiasing performance of FUDLR in cases where both popularity and implicit attribute biases exist in the ML1M dataset in Table \ref{tab:multibias}. Consistent with its performance in single-bias settings, FUDLR significantly outperforms the backbone model (BIGRec) across all metrics. Regarding recommendation utility, FUDLR achieves notable performance, improving HR@5 by 4.55\% and NDCG@5 by 5.60\%. This indicates that our method preserves and even enhances recommendation accuracy while mitigating bias. In terms of fairness, FUDLR demonstrates robust adaptability. It improves popularity fairness ($F_{\rm pop}@5$) by 3.42\% and attribute fairness ($F_{\rm attr}@5$) by 4.03\%. Additionally, it successfully reduces DP scores and improves APT scores, confirming its ability to handle multifaceted debiasing scenarios effectively. Furthermore, the runtime drops from 34.79 hours to 1.55 hours, representing a 95.53\% reduction in computational costs.

\begin{table*}[ht]
  \centering
  \caption{Performance Comparison for multiple co-existing bias mitigation in the ML1M dataset. The best results are in \textbf{bold}. The Improve is calculated as the relative improvement of our method over BIGRec. The \textcolor[rgb]{1,0,0}{red} Improve values indicate improved performance, while the \textcolor[rgb]{ 0,  .439,  .753}{blue} ones indicate degraded performance. *the improvement is significant at p<0.05.}
  \renewcommand{\arraystretch}{0.92}
  \vspace{-5pt}
  \resizebox{\linewidth}{!}{
    \begin{tabular}{c|cc|cc|cc|cc|cc|cc|c}
    \hline
    \multirow{2}[2]{*}{Methods} & \multicolumn{2}{c|}{HR$\uparrow$} & \multicolumn{2}{c|}{NDCG$\uparrow$} & \multicolumn{2}{c|}{APT$\uparrow$} & \multicolumn{2}{c|}{DP$\downarrow$} & \multicolumn{2}{c|}{$F_{\rm pop}\uparrow$} & \multicolumn{2}{c|}{$F_{\rm attr}\uparrow$} & \multirow{2}[2]{*}{Time (h) $\downarrow$} \\
          & K=5   & K=20  & K=5   & K=20  & K=5   & K=20  & K=5   & K=20  & K=5   & K=20  & K=5   & K=20  &  \\
    \hline
    BIGRec & 0.0220 & 0.0336 & 0.0167 & 0.0200 & 0.3256 & 0.2476 & 0.6361 & 0.5877 & 0.1644 & 0.2002 & 0.1947 & 0.2833 & 34.79 \\
    \hline
    FUDLR  & \textbf{0.0230*} & \textbf{0.0350*} & \textbf{0.0177*} & \textbf{0.0210*} & \textbf{0.3263*} & \textbf{0.2482*} & \textbf{0.6339*} & \textbf{0.5726*} & \textbf{0.1701*} & \textbf{0.2053*} & \textbf{0.2025*} & \textbf{0.2934*} & 1.55 \\
    
    \rowcolor[rgb]{ .851,  .851,  .851} \textit{Improve (\%)} & \textcolor[rgb]{ 1,  0,  0}{4.55} & \textcolor[rgb]{ 1,  0,  0}{4.17} & \textcolor[rgb]{ 1,  0,  0}{5.60} & \textcolor[rgb]{ 1,  0,  0}{5.15} & \textcolor[rgb]{ 1,  0,  0}{0.22} & \textcolor[rgb]{ 1,  0,  0}{0.23} & \textcolor[rgb]{ 1,  0,  0}{0.35} & \textcolor[rgb]{ 1,  0,  0}{2.57} & \textcolor[rgb]{ 1,  0,  0}{3.42} & \textcolor[rgb]{ 1,  0,  0}{2.54} & \textcolor[rgb]{ 1,  0,  0}{4.03} & \textcolor[rgb]{ 1,  0,  0}{3.58} &\textcolor[rgb]{ 1,0,0}{95.53} \\
    \hline
    \end{tabular}}
    \vspace{-5pt}
  \label{tab:multibias}
\end{table*}

\begin{table}[ht]
  \centering
  \caption{Ablation study in the ML1M dataset. The best results are highlighted in bold.}
  \vspace{-5pt}
  \renewcommand{\arraystretch}{0.95}
  \resizebox{\linewidth}{!}{
    \begin{tabular}{ccc|cc|cc|cc}
    \hline
    \multicolumn{3}{c|}{Variants} & \multicolumn{2}{c|}{HR$\uparrow$} & \multicolumn{2}{c|}{APT$\uparrow$} & \multicolumn{2}{c}{$F_{\rm pop}\uparrow$} \\
    Fair  & Acc   & Sparse & K=5   & K=20  & K=5   & K=20  & K=5   & K=20 \\
    \hline
    $\checkmark$     & -     & -     & 0.0184 & 0.0312 & \textbf{0.3698} & \textbf{0.2641} & 0.1473 & 0.1961 \\
    -     & $\checkmark$     & -     & 0.0204 & 0.0332 & 0.2818 & 0.2247 & 0.1498 & 0.1910 \\
    -     & -     & $\checkmark$     & 0.0220 & 0.0336 & 0.3256 & 0.2476 & 0.1644 & 0.2002 \\
    $\checkmark$     & $\checkmark$     & -     & 0.0202 & 0.0336 & 0.2854 & 0.2276 & 0.1492 & 0.1933 \\
    $\checkmark$     & -     & $\checkmark$     & 0.0208 & 0.0338 & 0.2836 & 0.2264 & 0.1522 & 0.1935 \\
    -     & $\checkmark$     & $\checkmark$     & 0.0206 & 0.0332 & 0.2798 & 0.2239 & 0.1506 & 0.1907 \\
    $\checkmark$     & $\checkmark$     & $\checkmark$     & \textbf{0.0226} & \textbf{0.0340} & 0.3278 & 0.2486 & \textbf{0.1681} & \textbf{0.2019} \\
    \hline
    \end{tabular}}
    \vspace{-5pt}
  \label{tab:ablation}
\end{table}

\subsection{RQ4: Model Analysis}
\subsubsection{Ablation Study}

To evaluate the individual contributions of the components within our bias identification module, we conducted an ablation study focusing on popularity debiasing in ML1M dataset. Specifically, we assess seven variants of FUDLR by permuting three core objectives in our mask optimization Eq. (\ref{eq:Overall_Objective}): accuracy preservation (Acc), fairness enhancement (Fair), and sparsity induction (Sparse). From Table \ref{tab:ablation}, we draw several key conclusions:

1) The Fair objective effectively targets bias but compromises accuracy. The variant with only the Fair objective significantly improves fairness metrics in the ML1M dataset, achieving the highest APT score. However, this improvement incurs a substantial cost to recommendation utility. For instance, HR@5 drops to 0.0184 from a baseline of 0.0220. This highlights the importance of balancing fairness and accuracy for practical recommendations.

2) The Acc objective is essential for maintaining utility. When the model is optimized solely using the Acc objective, it retains high recommendation accuracy (e.g., an HR@20 of 0.0332). However, it exhibits poor fairness performance, recorded as a low APT@20 of 0.2247. This suggests that a framework focused exclusively on accuracy fails to rectify the underlying data biases.

3) The full model yields the optimal performance balance. By integrating all three objectives, the FUDLR framework consistently achieves the highest accuracy alongside robust fairness. These results validate the efficacy of our method in achieving a harmonious balance between multiple objectives.

\begin{figure}[h]
    \centering
    \includegraphics[width=0.95\linewidth]{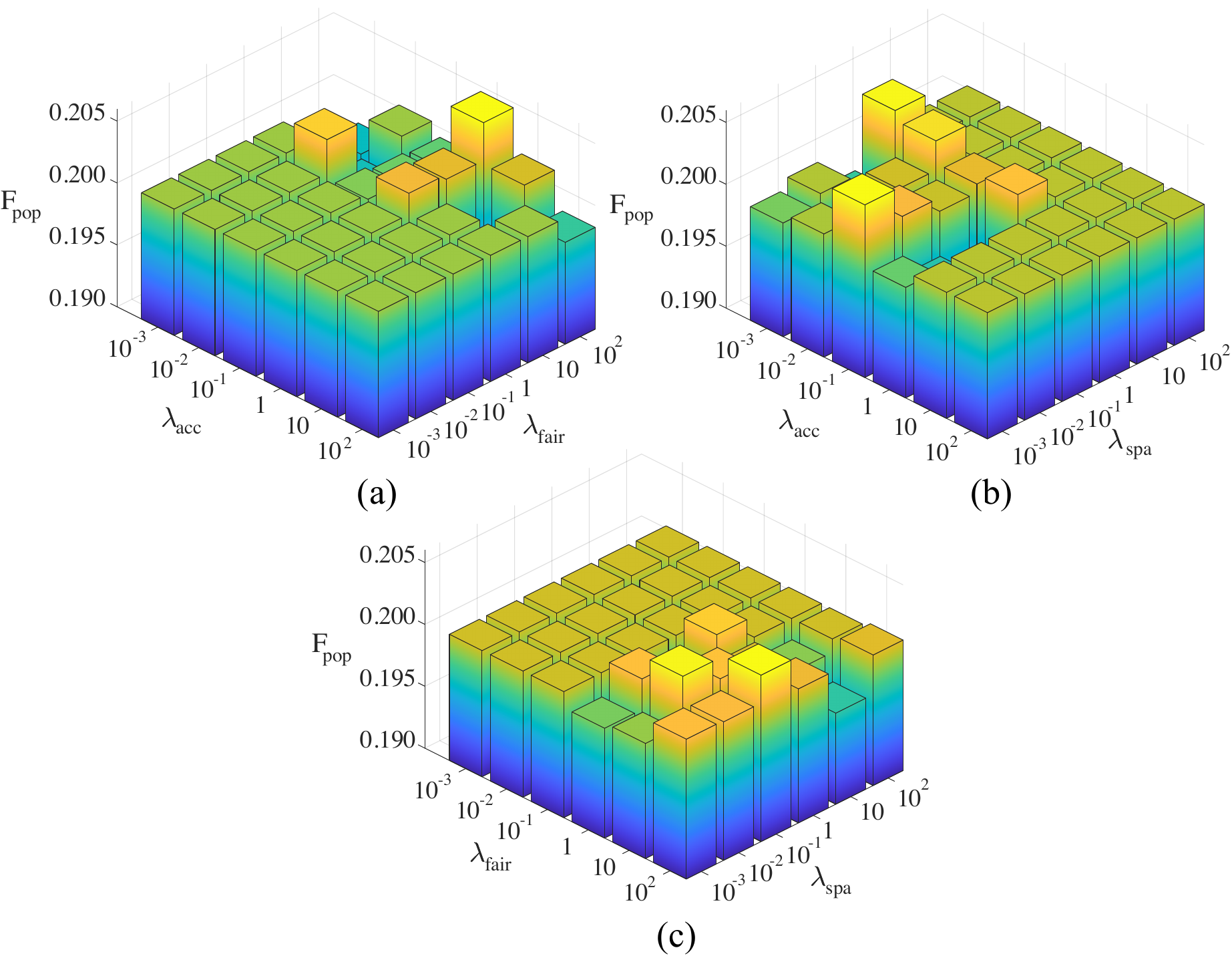}
    \vspace{-10pt}
    \caption{Impacts of weighting parameters in FUDLR on popularity debiasing performance {$F_{\rm pop}$} in the ML1M dataset.}
    \label{fig:three_graphs}
    \vspace{-12pt}
\end{figure}

\subsubsection{Parameter Analysis}
\label{sec:param_analysis}

We analyze the sensitivity of FUDLR to its key hyperparameters, {$\lambda_{\rm acc}$}, {$\lambda_{\rm fair}$}, and {$\lambda_{\rm spa}$}. Two hyperparameters are varied over {$\{10^i\}_{i=-3}^{2}$} while the third is fixed at its optimal value. Popularity bias and attribute bias results are shown in Figure \ref{fig:three_graphs} and Appendix \ref{appen:Parameter Analysis}, respectively. We observe the following patterns: \textbf{1) Accuracy and Fairness Trade-off:} Figure \ref{fig:three_graphs} (a) indicates the trade-off between accuracy and fairness. Setting extreme values of either parameter yields poor performance. Specifically, an excessively low {$\lambda_{\rm fair}$} provides insufficient debiasing, whereas an excessively high value removes informative interactions and harms accuracy. Conversely, a high {$\lambda_{\rm acc}$} enforces excessive conservatism and retains existing bias, while a very low value sacrifices too much recommendation utility. Empirical evidence suggests that moderate values yield the best performance. \textbf{2) Impact of Sparsity:} The sparsity weight regulates the magnitude of the unlearning process. As shown in \ref{fig:three_graphs} (b), setting high {$\lambda_{\rm spa}$} continuously degrades performance, as it drives the unlearning mask toward zero and causes the reversion to the backbone model. Conversely, a low {$\lambda_{\rm spa}$} relaxes constraints on interaction removal, making the outcome highly sensitive to the balance between {$\lambda_{\rm acc}$} and {$\lambda_{\rm fair}$}. Optimal results are typically observed in the moderate-to-high range, offering the targeted removal of a sparse set of influential interactions.

\section{Conclusion}
In this paper, we addressed two critical challenges in the realm of fair Large Language Model-based recommender systems: the limited generality of existing methods and the prohibitive cost of retraining. We proposed FUDLR, a general and efficient framework that reformulates debiasing through the lens of machine unlearning. Our two-stage approach first employs a novel bias identification module, utilizing a learnable mask optimized to balance fairness improvement and accuracy preservation. Notably, the bias-agnostic design of this module allows for seamless adaptation to various data biases by simply integrating the corresponding fairness metrics. Then, FUDLR performs efficient debiasing using influence-based unlearning techniques to remove the impact of identified biased data. Extensive experiments verify that FUDLR effectively mitigates various types of bias, including item-side popularity bias, user-side attribute bias, and complex co-existing biases. In the future, we plan to explore finer-grained and personalized debiasing in LLM-RS.

\begin{acks}
This work is partially supported by Australia ARC LP220100453 and ARC DP240100955. 
\end{acks}
\clearpage
\bibliographystyle{ACM-Reference-Format}
\bibliography{sample-base}

\appendix

\section{Proof Sketch of Proposition 1}
\label{appen:proof}
Here, we provide a brief proof sketch of Proposition \ref{prop:unlearning} based on the classical influence function technique \cite{DBLP:conf/icml/KohL17}.

Let the empirical risk be
\begin{equation}
    R(\theta) = \frac{1}{n} \sum_{z\in\mathcal{D}_{\rm train}} \mathcal{L}_{\rm LLM}(z; \theta),
\end{equation}
and denote its Hessian by $\mathbf{H}_\theta=\nabla^2_\theta R(\theta)$. By assuming the trained parameter $\theta$ is a stationary point of $R$, so $\nabla_\theta R(\theta)=0$.

Removing a single training example $z_k$ can be considered as down-weighting that example by $\epsilon=-\frac{1}{n}$. Thus, we obtain the following perturbed risk for unlearning a set $\mathcal{D}_{\rm unlearn}$:
\begin{equation}
    R_{\epsilon}(\theta) = R(\theta) + \epsilon \sum_{z_k\in\mathcal{D}_{\rm unlearn}} \mathcal{L}_{\rm LLM}(z_k; \theta).
\end{equation}
Let $\theta+\Delta\theta$ be the minimizer of $R_\epsilon$. The stationary condition for the perturbed risk is:
\begin{equation}
    \nabla_\theta R_\epsilon(\theta+\Delta\theta) = \nabla_\theta R(\theta+\Delta\theta) + \epsilon \sum_{z_k\in\mathcal{D}_{\rm unlearn}} \nabla_\theta \mathcal{L}_{\rm LLM}(z_k; \theta+\Delta\theta) = 0.
\end{equation}
We perform a first-order Taylor expansion, and since $\nabla_\theta R(\theta)=0$, we have $\nabla_\theta R(\theta+\Delta\theta)\approx\mathbf{H}_\theta\Delta\theta$. Thus, we get
\begin{equation}
    \mathbf{H}_\theta\Delta\theta + \epsilon \sum_{z_k\in\mathcal{D}_{\rm unlearn}} \nabla_\theta \mathcal{L}_{\rm LLM}(z_k; \theta) \approx 0.
\end{equation}
Solving for $\Delta\theta$ gives
\begin{equation}
    \Delta\theta \approx \frac{1}{n} \mathbf{H}_\theta^{-1} \sum_{z_k\in\mathcal{D}_{\rm unlearn}} \nabla_\theta \mathcal{L}_{\rm LLM}(z_k; \theta).
\end{equation}
This completes the proof sketch of Proposition \ref{prop:unlearning}.

Note that, since the parameters $\theta$ of LLMs may be obtained in a non-convex setting and may therefore correspond only to a local optimum, we refer to recent studies \cite{DBLP:conf/icml/KohL17,DBLP:conf/nips/ChenYXBHHFZWL23} that investigate the reliability of influence functions under such conditions.

\section{Datasets}
\label{appen:Datasets details}
Table \ref{tab:dataset_stats} summarizes the statistics of the datasets employed in our experiments. The popular benchmark dataset, ML1M, comprises 1,000,209 ratings from 6,040 users and 3,952 movies. It features abundant user demographic attributes, while its low density (0.04190) suggests a long-tailed item popularity distribution. Games, a subset of the Amazon Review dataset, spans 55,223 users and 17,408 items. Its low density (0.00052) indicates severe sparsity and significant popularity bias.

\begin{table}[ht]
      \vspace{-5pt}
    \centering
    \caption{Statistics of the datasets used in the experiments.}
    \label{tab:dataset_stats}

    \resizebox{\columnwidth}{!}{%
        \begin{tabular}{c c c c c c}                    
            \toprule
            \textbf{Dataset} & \textbf{\#Users} & \textbf{\#Items} & \textbf{\#Interactions} & \textbf{Density} & \textbf{Bias Type} \\
            \midrule
            ML1M   & 6,040   & 3,952  & 1,000,209 & 0.04190 & Popularity Bias, Attribute Bias \\
            Games & 55,233 & 17,408 & 497,577 & 0.00052 & Popularity Bias \\
            \bottomrule
        \end{tabular}}%
\end{table}

\section{Parameter Analysis}
\label{appen:Parameter Analysis}
In this section, we analyze the sensitivity of FUDLR with respect to attribute debiasing performance. Two hyperparameters (randomly selected from $\lambda_{\rm acc}$}, {$\lambda_{\rm fair}$}, and {$\lambda_{\rm spa}$}) are varied over {$\{10^i\}_{i=-3}^{2}$} while the third is fixed at its optimal value. From Figure \ref{fig:param_ml1m_attr}, we make the following observations: \textbf{1) Balancing between Accuracy and Fairness:} Figure \ref{fig:param_ml1m_attr} (a) underscores the fundamental trade-off between accuracy and fairness, driven by {$\lambda_{\rm acc}$} and {$\lambda_{\rm fair}$}. Excessive values of either parameter cause poor effectiveness: an extremely small {$\lambda_{\rm fair}$} leads to insufficient debiasing, whereas an extremely large value removes informative interactions and undermines accuracy. Conversely, a high {$\lambda_{\rm acc}$} enforces excessive conservatism, effectively maintaining existing bias, while a very low value sacrifices too much recommendation utility. Empirical evidence suggests that moderate values (e.g., $10^{-1}$ to 1) result in the best performance, emphasizing the necessity of trading-off these two objectives. \textbf{2) Impact of Sparsity:} The sparsity weight, {$\lambda_{\rm spa}$}, regulates the magnitude of the unlearning process. As shown in \ref{fig:param_ml1m_attr} (c), setting an excessive {$\lambda_{\rm spa}$} consistently degrades performance. A stringent sparsity constraint drives the unlearning mask toward zero, effectively preventing the modification of interactions and causing the model to remain in its initial biased state. In contrast, a low {$\lambda_{\rm spa}$} relaxes constraints on interaction removal, making the outcome highly sensitive to the balance between {$\lambda_{\rm acc}$} and {$\lambda_{\rm fair}$}. Optimal results are observed in the moderate-to-high range (typically [$10^{-1}$, 10]), which encourages the targeted removal of a sparse yet significant set of interactions.

\begin{figure}[htbp]
    \centering
    \includegraphics[width=\linewidth]{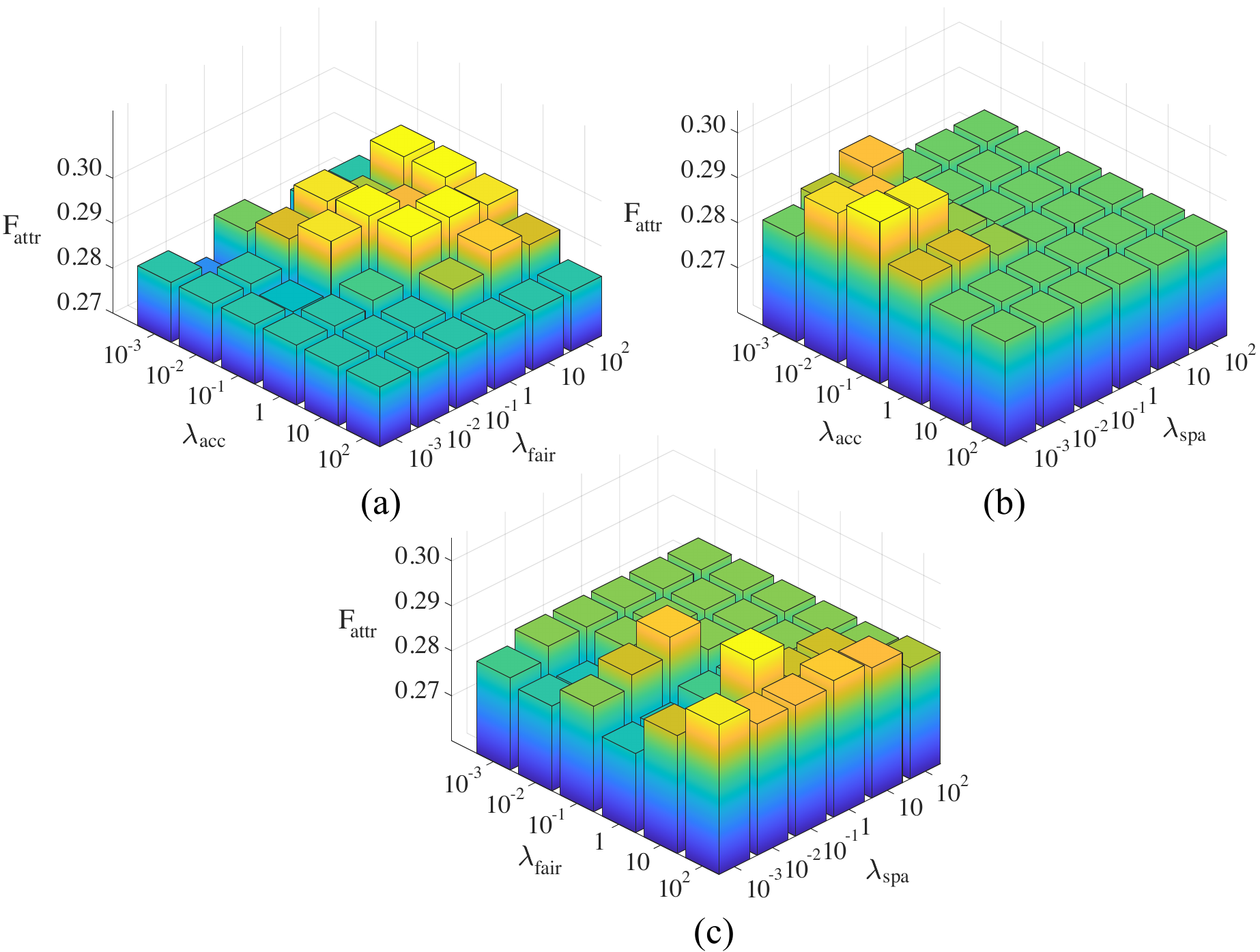}
    \caption{Impacts of weighting parameters in FUDLR on attribute debiasing performance {$F_{\rm attr}$} in the ML1M dataset.}

    \label{fig:param_ml1m_attr}
    \vspace{-12pt}
\end{figure}

\renewcommand{\thesection}{\Alph{section}}

\section{Case Study}
\label{appen:case_study}
To illustrate how FUDLR mitigates popularity bias and attribute bias, we present case studies (Table \ref{tab:case_study1_1}, Table \ref{tab:case_study1_2}, Table \ref{tab:case_study2}) using randomly selected samples from both ML1M and Games datasets. Our results indicate FUDLR achieves the best performance in terms of both bias alleviation and accuracy preservation.

\begin{table}[ht]
    \vspace{-5pt}
    \centering

    \caption{A case study on popularity bias in the ML1M and Games datasets.}
    \label{tab:case_study1_1}
    
    \scriptsize 
    \renewcommand{\arraystretch}{2} 
    \setlength{\tabcolsep}{0.6pt}
    
    \resizebox{\columnwidth}{!}{%
        \begin{tabular}{c|c|c| c |c| c | c}
            \hline
            \rowcolor{white} 
            \textbf{User} & \textbf{Dataset} & \textbf{Methods} & \textbf{Recommended Items} & \textbf{Popular Status} & \textbf{Debiased} & \textbf{Targeted} \\
            \hline
            
            \multirow{4}{*}[-1em]{User\_1} & \multirow{4}{*}[-1em]{ML1M} 
              & \textbf{Ground Truth} & \textbf{``Omen, The (1976)''} & \textbf{\textcolor{blue}{Unpopular}} & \textbf{--} & \textbf{--} \\ \cline{3-7} 
              
              & & BIGRec & ``Superman III (1983)'' & \textcolor{red}{Popular} & \xmark & \xmark \\ \cline{3-7}
              
              & & RWRR & ``Til There Was You (1997)'' & \textcolor{blue}{Unpopular} & \cmark & \xmark \\ \cline{3-7}
              
              & & \cellcolor{gray!20}FUDLR & \cellcolor{gray!20}\textbf{``Omen, The (1976)''} & \cellcolor{gray!20}\textcolor{blue}{Unpopular} & \cellcolor{gray!20}\cmark & \cellcolor{gray!20}\cmark \\
            \hline
    
            \multirow{4}{*}[-1em]{User\_2} & \multirow{4}{*}[-1em]{Games} 
              & \textbf{Ground Truth} & \textbf{``eXtremeRate\&reg; Textured Red Back Panels......''} & \textbf{\textcolor{blue}{Unpopular}} & \textbf{--} & \textbf{--} \\ \cline{3-7}
              
              & & BIGRec & ``Generic Battery Pack Cover for Xbox 360 Controller'' & \textcolor{red}{Popular} & \xmark & \xmark \\ \cline{3-7}
              
              & & RWRR & ``Disney Infinity 3.0 Edition: Star Wars Darth Vader Figure'' & \textcolor{blue}{Unpopular} & \cmark & \xmark \\ \cline{3-7}
              
              & & \cellcolor{gray!20}FUDLR & \cellcolor{gray!20} \textbf{``eXtremeRate\&reg; Textured Red Back Panels......''} & \cellcolor{gray!20}\textcolor{blue}{Unpopular} & \cellcolor{gray!20}\cmark & \cellcolor{gray!20}\cmark \\
            \hline
            
        \end{tabular}}%
\end{table}

\begin{table}[t]
    \centering    
    \vspace{-5pt}
    \caption{A case study on popularity bias in implicit and explicit scenarios from the ML1M dataset.}

    \label{tab:case_study1_2}
    
    \scriptsize 
    \renewcommand{\arraystretch}{2} 
    \setlength{\tabcolsep}{0.6pt}
    
    \resizebox{\columnwidth}{!}{%
        \begin{tabular}{c|c|c| c |c| c | c}
            \hline
            \rowcolor{white} 
            \textbf{User} & \textbf{Scenarios} & \textbf{Methods} & \textbf{Recommended Items} & \textbf{Popular Status} & \textbf{Debiased} & \textbf{Targeted} \\
            \hline
            
            \multirow{4}{*}[-1em]{User\_3} & \multirow{4}{*}[-1em]{Implicit} 
              & \textbf{Ground Truth} & \textbf{``Force 10 from Navarone (1978)''} & \textbf{\textcolor{blue}{Unpopular}} & \textbf{--} & \textbf{--} \\ \cline{3-7} 
              
              & & BIGRec & ``Evil Dead II (Dead By Dawn) (1987)'' & \textcolor{red}{Popular} & \xmark & \xmark \\ \cline{3-7}
              
              & & CFP & ``Return to Me (2000)'' & \textcolor{blue}{Unpopular} & \cmark & \xmark \\ \cline{3-7}
              
              & & \cellcolor{gray!20}FUDLR & \cellcolor{gray!20}\textbf{``Force 10 from Navarone (1978)''} & \cellcolor{gray!20}\textcolor{blue}{Unpopular} & \cellcolor{gray!20}\cmark & \cellcolor{gray!20}\cmark \\
            \hline
    
            \multirow{4}{*}[-1em]{User\_4} & \multirow{4}{*}[-1em]{Explicit} 
              & \textbf{Ground Truth} & \textbf{``Home Alone 2: Lost in New York (1992)''} & \textbf{\textcolor{blue}{Unpopular}} & \textbf{--} & \textbf{--} \\ \cline{3-7}
              
              & & BIGRec & ``Rocky Horror Picture Show, The (1975)'' & \textcolor{red}{Popular} & \xmark & \xmark \\ \cline{3-7}
              
              & & Masking & ``Home Alone 3 (1997)'' & \textcolor{blue}{Unpopular} & \cmark & \xmark \\ \cline{3-7}
              
              & & \cellcolor{gray!20}FUDLR & \cellcolor{gray!20} \textbf{``Home Alone 2: Lost in New York (1992)''} & \cellcolor{gray!20}\textcolor{blue}{Unpopular} & \cellcolor{gray!20}\cmark & \cellcolor{gray!20}\cmark \\
            \hline
            
        \end{tabular}}%
\end{table}

\begin{table}[t]
    \centering    
    \caption{A case study on attribute bias in the ML1M dataset.}
    \vspace{-5pt}
    
    \label{tab:case_study2}
    
    \scriptsize 
    \renewcommand{\arraystretch}{2} 
    \setlength{\tabcolsep}{0.6pt}
    
    \resizebox{\columnwidth}{!}{%

        \begin{tabular}{c|c|c| c |c| c}
            \hline
            \rowcolor{white} 
            \textbf{User} & \textbf{Gender} & \textbf{Methods} & \textbf{Recommended Items} & \textbf{Genres} & \textbf{Targeted} \\
            \hline
            
            \multirow{4}{*}[-1em]{User\_1} & \multirow{4}{*}[-1em]{Male} 
              & \textbf{Ground Truth} & \textbf{``Goodbye Girl, The (1977)''} & \textbf{{Romance and Comedy}} & \textbf{--} \\ \cline{3-6} 
              
              & & BIGRec & ``20,000 Leagues Under the Sea (1954)'' & {War and Action} & \xmark \\ \cline{3-6}
              
              & & Masking & ``Star Trek: The Wrath of Khan (1982)'' & {Science Fiction and Adventure} & \xmark \\ \cline{3-6}
              
              & & \cellcolor{gray!20}FUDLR & \cellcolor{gray!20}\textbf{``Goodbye Girl, The (1977)''} & \cellcolor{gray!20} \textbf{Romance and Comedy} & \cellcolor{gray!20}\cmark \\
            \hline
    
            \multirow{4}{*}[-1em]{User\_2} & \multirow{4}{*}[-1em]{Female} 
              & \textbf{Ground Truth} & \textbf{``Batman (1989)''} & \textbf{Action and Adventure} & \textbf{--} \\ \cline{3-6}
              
              & & BIGRec & ``Til There Was You'' & {Romance and Comedy} & \xmark \\ \cline{3-6}
              
              & & Masking & ``Til There Was You'' & {Romance and Comedy} & \xmark \\ \cline{3-6}
              
              & & \cellcolor{gray!20}FUDLR & \cellcolor{gray!20} \textbf{``Batman (1989)''} & \cellcolor{gray!20}\textbf{Action and Adventure} & \cellcolor{gray!20}\cmark \\
            \hline
            
        \end{tabular}}%
\end{table}
\subsection{Popularity Bias}
We randomly select a sample from the ML1M and Games datasets,  respectively, to evaluate the effectiveness of FUDLR in popularity bias alleviation. As shown in Table \ref{tab:case_study1_1}, FUDLR demonstrates a superior competence to strike the balance between fairness and accuracy in both datasets. For instance, in the ML1M dataset, when the interaction item (Ground Truth) for User\_1 is the unpopular item ``Omen, The (1976)'', the backbone method (BIGRec) fails by incorrectly recommending the popular item ``Superman III (1983)''. While another baseline methodology (RWRR) successfully eliminates popularity bias, it is not capable of targeting the exact item.  

Regarding the popular bias in both implicit and explicit contexts within the ML1M dataset (Table \ref{tab:case_study1_2}), FUDLR consistently indicates remarkable performance in trading off fairness and accuracy. Overall, both cases highlight the superior effectiveness of FUDLR to alleviate popularity bias without compromising accuracy.

\subsection{Attribute Bias}
A random sample from the ML1M dataset is selected for each sensitive attribute (e.g., gender) to examine the capability of FUDLR in attribute bias mitigation. As displayed in Table \ref{tab:case_study2}, FUDLR demonstrates significant effectiveness in eliminating attribute bias, while matching users' genuine preferences, outperforming all other approaches. For example, User\_2, who is a female, shows an inclination toward ``Batman (1989)'', an Action and Adventure movie. However, baseline models mistakenly recommend Romance and Comedy films, likely reflecting gender-based stereotypes. In contrast, FUDLR achieves a perfect balance between accuracy and fairness. This case illustrates FUDLR’s strong ability to mitigate attribute bias while maintaining high recommendation accuracy.

\section{Unlearning Estimation Analysis}
\label{appen:estimation}
Since the proposed FUDLR aims to perform efficient debiasing by estimating and mitigating the influence of identified samples, it is important to verify whether the influence estimation and update truly reflect the target—namely, retraining the model with $\mathcal{D}_{\rm remain}$. We verify this by comparing FUDLR with the retrained model for both popularity and attribute debiasing on the ML1M dataset. The performance difference (``GAP'') is reported in Table \ref{tab:retraining_GAP}, which shows a negligible gap in both accuracy and fairness scores. Specifically, the difference in accuracy is minimal, with gaps of only 0.0026 for HR@20 and 0.0029 for NDCG@20. Similarly, the disparities for {$F_{\rm pop}$} (K=20) and {$F_{\rm attr}$} (K=5) are approximately 0.0126 and 0.0440. For debiasing effectiveness at K=5, the gaps for APT and DP are as low as 0.0073 and 0.0022. Furthermore, the runtime is reduced by 31.70 hours, demonstrating that FUDLR is significantly more efficient.

Overall, FUDLR achieves near-optimal performance compared to the retrained unbiased model. This highlights that our method can effectively and efficiently approximate the retrained model with high fidelity for bias mitigation.

\begin{table}[ht]
  \centering
  \renewcommand{\arraystretch}{1.8} 
  \setlength{\tabcolsep}{0.6pt}
  
  \caption{Performance GAP between FUDLR and Retrained Unbiased Model for multifaceted bias mitigation in the ML1M dataset.}
  \vspace{-5pt}
  
  \resizebox{\linewidth}{!}{
    \begin{tabular}{c|cc|cc|cc|cc|cc|cc|c}
    \hline
    \multirow{2}[2]{*}{} & \multicolumn{2}{c|}{HR} & \multicolumn{2}{c|}{NDCG} & \multicolumn{2}{c|}{APT} & \multicolumn{2}{c|}{DP} & \multicolumn{2}{c|}{$F_{\rm pop}$} & \multicolumn{2}{c|}{$F_{\rm attr}$} & \multirow{2}[2]{*}{Time (h) } \\
          & K=5   & K=20  & K=5   & K=20  & K=5   & K=20  & K=5   & K=20  & K=5   & K=20  & K=5   & K=20  &  \\
    \hline
    
    \rowcolor[rgb]{ .851,  .851,  .851} 
    \textbf{GAP} & 0.0028 & 0.0026 & 0.0030 & 0.0029 & 0.0073 & 0.0105 & 0.0022 & 0.0181 & 0.0166 & 0.0126 & 0.0440 & 0.0613 & 31.70 \\
    \hline
    \end{tabular}%
  }
  \vspace{-5pt}
  \label{tab:retraining_GAP}
\end{table}

\end{document}